\documentclass[utf8]{frontiersFPHY} 

\setcitestyle{square} 
\usepackage{url,hyperref,lineno,microtype,subcaption}
\usepackage[onehalfspacing]{setspace}
\usepackage{braket}
\usepackage{multirow}

\def\keyFont{\fontsize{8}{11}\helveticabold }
\def\firstAuthorLast{M. Shiraishi} 
\def\Authors{Maresuke Shiraishi\,$^{1,*}$}
\def\Address{$^{1}$Department of General Education, National Institute of Technology, Kagawa College, 355 Chokushi-cho, Takamatsu, Kagawa 761-8058, Japan}
\def\corrAuthor{Corresponding Author}
\def\corrEmail{shiraishi-m@t.kagawa-nct.ac.jp}

\begin{document}
\onecolumn
\firstpage{1}



\title[Tensor Non-Gaussianity Search]{Tensor Non-Gaussianity Search: Current Status and Future Prospects}

\author[\firstAuthorLast ]{\Authors} 
\address{} 
\correspondance{} 

\extraAuth{}

\maketitle

\begin{abstract}


  Primordial gravitational waves (GWs) are said to be a smoking gun in cosmic inflation, while, even if they are detected, the specification of their origins are still required for establishing a true inflationary model. Testing non-Gaussianity in the tensor-mode anisotropies of the cosmic microwave background (CMB) is one of the most powerful ways to identify sources of GW signals. In this paper, we review studies searching for tensor non-Gaussianities employing the CMB bispectrum and forecast future developments. No significant signal has so far been found from temperature and E-mode polarization data, while orders-of-magnitude improvements in detection limits can be achieved by adding the information of B-mode polarization. There is already an established methodology for bispectrum estimation, which encourages a follow-up investigation with next-decadal CMB B-mode surveys.

\tiny
 \keyFont{ \section{Keywords:} CMB, non-Gaussianity, gravitational wave, inflation, observational constraints, B-mode} 
\end{abstract}

\section{Introduction}

Recent cosmic-variance-limited-level measurements of the scalar sector in primordial fluctuations using the temperature and $E$-mode polarization fields of the cosmic microwave background (CMB), indicate that the Universe experienced an inflationary expansion at very early stages \citep{Aghanim:2018eyx,Akrami:2018odb}. Now, compelling evidence is expected to lie in the tensor sector; namely, the primordial gravitational wave (GW). Even without any particular source, primordial GWs arise naturally from inflationary vacuum fluctuations. Since their amplitude directly reflects the inflationary energy scale, various observational projects aim at hunting them. There, the measurements of the large-scale CMB $B$-mode polarization have attracted everyone's attention as it is a distinctive observable of the tensor mode (see e.g., Refs.~\citep{Kamionkowski:2015yta,Guzzetti:2016mkm} and references therein). 

In order to investigate the primordial tensor sector using the CMB anisotropies, the power spectrum (2-point correlation) is primarily employed. In single-field inflation with Einstein gravity, GWs are nearly Gaussian and hence the higher-order correlations vanish \citep{Acquaviva:2002ud,Maldacena:2002vr}. In this case, only the power spectrum becomes an informative observable. This paper, however, will address the potential of the bispectrum (3-point correlation) as a probe of inflationary physics.

One possible way, making it more informative, is the addition of some extra source fields. In the presence of the gauge field, the anisotropic stress fluctuations are formed, generating GWs. Because this is a non-linear process, the resulting GWs are non-Gaussian (NG) and therefore a large tensor bispectrum can be realized. In particular, the production of NG GWs becomes more efficient when the gauge field is coupled to the axion \citep{Namba:2015gja,Agrawal:2017awz}. Moreover, production due to higher-spinning particles has been studied intensively in recent years \citep{Dimastrogiovanni:2018gkl,Goon:2018fyu}. There is also the possibility of a post-inflationary generation of NG GWs due to magnetic fields \citep{Shiraishi:2011dh}. Another possible way to produce large tensor NGs comes from non-trivial non-linear gravitational interactions predicted in some modified theories of gravity (e.g. \citep{Maldacena:2011nz,Gao:2011vs,Gao:2012ib,Akita:2015mho,Domenech:2017kno,Bartolo:2017szm,Naskar:2018rmu,Anninos:2019nib,Ozsoy:2019slf}). Then, the induced bispectra can have distinctive shapes compared with that from the usual Einstein term. 

In this sense, the CMB tensor-mode bispectrum is a key indicator of the inflationary particle content and/or high energy gravity. Since a general formalism was established in Refs.~\citep{Shiraishi:2010sm,Shiraishi:2010kd}, various signatures of various theoretical models in the CMB bispectra have been studied (see Refs.~\citep{Shiraishi:2011dh,Shiraishi:2012rm,Shiraishi:2012sn,Shiraishi:2013vha} for primordial magnetic field (PMF) models, Refs.~\citep{Shiraishi:2013kxa,Namba:2015gja,Shiraishi:2016yun} for the axion inflation ones, Refs.~\citep{Shiraishi:2010kd,Shiraishi:2011st,Domenech:2017kno,Tahara:2017wud,Bartolo:2018elp} for modified gravity ones and Refs.~\citep{Meerburg:2016ecv,Kothari:2019yyw} for other motivations). One of the most interesting phenomena discovered there (unseen in the scalar bispectrum analysis) is that the tensor bispectrum can yield the non-vanishing signal in not only even but also odd $\ell_1 + \ell_2 + \ell_3$ multipoles. The GW bispectra generate the CMB auto and cross bispectra including the $B$-mode polarization. In the usual theories where the GW bispectra are parity-invariant, the odd $\ell_1 + \ell_2 + \ell_3$ signal can arise in $TTB$, $TEB$, $EEB$, and $BBB$ \citep{Shiraishi:2013vha,Meerburg:2016ecv,Domenech:2017kno,Tahara:2017wud}. Moreover, parity-odd GW bispectra motivated by parity-breaking theories can also induce the non-zero odd $\ell_1 + \ell_2 + \ell_3$ signal in $TTT$, $TTE$, $TEE$, $TBB$, $EEE$, and $EBB$ \citep{Shiraishi:2011st,Shiraishi:2012sn,Shiraishi:2013kxa,Shiraishi:2016ads,Shiraishi:2016yun,Bartolo:2018elp}.

Now, any type of theoretical bispectrum template is testable with the temperature and E/B-mode polarization data by use of a general bispectrum estimator for the even \citep{Fergusson:2009nv,Fergusson:2010dm,Fergusson:2014gea,Shiraishi:2019exr} and odd $\ell_1 + \ell_2 + \ell_3$ domain \citep{Shiraishi:2014roa,Shiraishi:2019exr}. The magnitudes of some representative tensor NG templates have already been estimated from the temperature map of WMAP \citep{Shiraishi:2013wua,Shiraishi:2014ila,Shiraishi:2017yrq} and the temperature and $E$-mode polarization from {\it Planck} \citep{Ade:2015cva,Ade:2015ava,Akrami:2019izv}. Absence of NG reported there constrains the various inflationary models mentioned above.

In this paper, besides reviewing previous studies, we discuss future prospects of the tensor NG search assuming the detection of $B$-mode polarization in next-generation CMB experiments. We found that, in terms of observational limits on the size of some scale-invariant templates, an improvement up to three orders of magnitude is expected assuming LiteBIRD-level sensitivities \citep{Hazumi:2012aa,Matsumura:2013aja,2016JLTP..tmp..169M}. 

This paper is organized as follows. In the next section, we summarize theoretical scenarios predicting characteristic primordial tensor bispectra. In Sec.~\ref{sec:CMB_bis}, we compute the CMB bispectra and show their behavior. In Sec.~\ref{sec:data}, we present the current observational constraints on tensor NGs and discuss future prospects assuming $B$-mode polarization surveys. The final section is devoted to the conclusions of this paper.

\section{Theoretical motivations} \label{sec:models}

In this section, we briefly review the primordial tensor bispectra predicted in some theoretical scenarios. For convenience, let us work with the helicity basis; thus, the primordial perturbation is represented as $\xi_{\bf k}^{(\lambda)}$ \citep{Shiraishi:2010kd}. Here, $\lambda$ denotes the helicity and takes $0$, $\pm 1$, and $\pm 2$ for the scalar, vector, and tensor modes, respectively. In practice, we identify $\xi_{\bf k}^{(0)}$ and $\xi_{\bf k}^{(\pm 2)}$ with the curvature perturbation $\zeta_{{\bf k}}$ and the GW $h_{{\bf k}}^{(\pm 2)}$, respectively.

Assuming statistical homogeneity, the primordial bispectrum takes the form
\begin{equation}
  \Braket{\prod_{n=1}^3 \xi_{{\bf k}_n}^{(\lambda_n)}}
  = (2\pi)^3 \delta^{(3)}\left(\sum_{n = 1}^3 {\bf k}_n\right) \Braket{\prod_{n=1}^3 \xi_{{\bf k}_n}^{(\lambda_n)}}' .
\end{equation}
  Then, let us split this in two
\begin{equation}
  \Braket{\prod_{n=1}^3 \xi_{{\bf k}_n}^{(\lambda_n)}}'
  = \mathcal{B}_{k_1 k_2 k_3}^{\lambda_1 \lambda_2 \lambda_3} A_{\hat{k}_1 \hat{k}_2 \hat{k}_3}^{\lambda_1 \lambda_2 \lambda_3} . 
\end{equation}
In what follows, how the shape, represented by $\mathcal{B}_{k_1  k_2 k_3}^{\lambda_1 \lambda_2 \lambda_3}$, and the angular structure, represented by $A_{\hat{k}_1 \hat{k}_2 \hat{k}_3}^{\lambda_1 \lambda_2 \lambda_3}$, changes depending on the model is argued.

\subsection{Extra sources}

Efficient tensor NG production is realized by adding extra source fields to the theory. In what follows, we discuss the vector field as a source.

\subsubsection{Inflationary axion-gauge coupling} \label{subsubsec:pseudo}

A characteristic tensor NG is realized in inflationary models involving the coupling between the axion $\phi$ and a gauge field like $f(\phi) \widetilde{F}F$, with $F$ and $\widetilde{F}$ the gauge field strength tensor and its dual, respectively \citep{Cook:2013xea,Namba:2015gja,Maleknejad:2016qjz,Dimastrogiovanni:2016fuu,Agrawal:2017awz,Agrawal:2018mrg}. In this case, the chirality of the gauge field is transferred into the GW sector and hence only the plus mode of GW, $h^{(+2)}$, survives. The resulting production is efficient at specific scales, and the tensor-tensor-tensor bispectrum peaks for equilateral configurations, $k_1 \sim k_2 \sim k_3$ \citep{Cook:2013xea}.

The detailed spectral feature varies with the shape of the coupling $f(\phi)$ or the axion potential \citep{Namba:2015gja,Agrawal:2018mrg}. For example, adopting $f(\phi) \propto \phi$ and a nearly flat potential, one can obtain a scale-invariant shape as \citep{Shiraishi:2013kxa}
\begin{eqnarray}
  \mathcal{B}_{k_1 k_2 k_3}^{\lambda_1 \lambda_2 \lambda_3}
  &=& \frac{16\sqrt{2}}{27} f_{\rm NL}^{ttt, \rm eq} 
  S_{k_1 k_2 k_3}^{\rm eq} \delta_{\lambda_1, 2} \delta_{\lambda_2, 2} \delta_{\lambda_3, 2} , \label{eq:B_pseudo} \\
   A_{\hat{k}_1 \hat{k}_2 \hat{k}_3}^{\lambda_1 \lambda_2 \lambda_3}
 &=& 
 e_{ij}^{(-\lambda_1)}(\hat{k}_1)
  e_{jk}^{(-\lambda_2)}(\hat{k}_2)
  e_{ki}^{(-\lambda_3)}(\hat{k}_3),
  \label{eq:A_pseudo}
\end{eqnarray}
where $e_{ij}^{(\lambda)}(\hat{k})$ is the polarization tensor defined in the helicity basis obeying $e_{ii}^{(\lambda)}(\hat{k}) = \hat{k}_i e_{ij}^{(\lambda)}(\hat{k}) = 0$, $e_{ij}^{(\lambda) *}(\hat{k}) = e_{ij}^{(-\lambda)}(\hat{k}) = e_{ij}^{(\lambda)}(- \hat{k})$, and $e_{ij}^{(\lambda)}(\hat{k}) e_{ij}^{(\lambda')}(\hat{k}) = 2 \delta_{\lambda, -\lambda'}$ \citep{Shiraishi:2010kd}, and $S_{k_1 k_2 k_3}^{\rm eq}$ is the usual scalar equilateral bispectrum template, given by:
\begin{equation}
  S_{k_1 k_2 k_3}^{\rm eq} \equiv \frac{18}{5}
  \left( 2\pi^2 \mathcal{P}_\zeta \right)^2  
  \left[ - \left(\frac{1}{k_1^{3} k_2^{3}} + 2 \ {\rm perm} \right)
    - \frac{2}{k_1^{2} k_2^{2} k_3^{2}}
    + \left( \frac{1}{k_1 k_2^{2} k_3^{3}} + 5 \ {\rm perm} \right) \right] .
\end{equation}
Here $\mathcal{P}_\zeta$ is the amplitude of the curvature power spectrum. A newly introduced parameter $f_{\rm NL}^{ttt, \rm eq}$ quantifies the relative size of the $\lambda = +2$ bispectrum to $S_{k_1 k_2 k_3}^{\rm eq}$ and satisfies%
\footnote{This $f_{\rm NL}^{ttt, \rm eq}$ is equivalent to $f_{\rm NL}^P$ in Ref.~\citep{Shiraishi:2014ila} and $f_{\rm NL}^{\rm tens}$ in Ref.~\citep{Ade:2015ava,Akrami:2019izv}.
}
\begin{equation}
  f_{\rm NL}^{ttt, \rm eq} \equiv \lim_{k_i \to k}
  \frac{\mathcal{B}_{k_1 k_2 k_3}^{+2+2+2} A_{\hat{k}_1 \hat{k}_2 \hat{k}_3}^{+2 +2 +2} }
       {S_{k_1 k_2 k_3}^{\rm eq} } . \label{eq:fnl_def_ttt_eq} 
\end{equation}

The expected size of the GW bispectrum, of course, depends on the case. If the axion is identified with the inflaton field, the amount of scalar production exceeds the GW one \citep{Barnaby:2011vw}. Then, the scalar bispectrum measurement gives much tighter constraints on the model \citep{Ade:2015ava,Akrami:2019izv}. In contrast, in multifield models where the inflaton field and the axion coexist, more effective GW production occurs. In the model where the axion is coupled to the U(1) gauge field, a characteristic bump appears in the GW bispectrum, and, depending on its location, it is detectable by the CMB $BBB$ bispectrum measurement \citep{Namba:2015gja,Shiraishi:2016yun}. However, at the same time, a similar bump in the GW power spectrum is measured with a higher signal-to-noise ratio from the CMB $BB$ power spectrum. The strongest constraints are obtained through the GW power spectrum measurement, and the GW bispectrum provides complementary information. In the model including a SU(2)-gauge coupling, the GW bispectrum can dominate the scalar one, and more interestingly, in some regions of the parameter space, the GW bispectrum has high detectability compared to the GW power spectrum \citep{Agrawal:2017awz,Agrawal:2018mrg}. The GW bispectrum is nearly scale-invariant and therefore parametrized by $f_{\rm NL}^{ttt, \rm eq}$, which is related to the tensor-to-scalar ratio $r$ and the energy density fraction of the SU(2) gauge field $\Omega_A$ according to \citep{Agrawal:2017awz}
  \begin{equation}
    f_{\rm NL}^{ttt, \rm eq} \sim 2.5 \frac{r^2}{\Omega_A} . \label{eq:fnl_ttt_eq_pseudo}
  \end{equation}
  This indicates that $f_{\rm NL}^{ttt, \rm eq}$ can take detectably large values, i.e., $f_{\rm NL}^{ttt, \rm eq} > 1$, when $\Omega_A$ is smaller than $r^2$.

\subsubsection{Primordial magnetic fields} \label{subsubsec:PMF}

If PMFs, $B_i$, are created via some non-trivial mechanism before inflation, they are stretched beyond the horizon by the inflationary expansion and there remain relic fields at large scales. Such a scenario has been widely argued to explain galactic or extragalactic magnetic fields observed at present (for reviews see Refs.~\citep{Widrow:2002ud,Kulsrud:2007an}). In this case, PMFs form the anisotropic stress fluctuations, and they source the GW at superhorizon scales until neutrino decoupling. The induced GW is called a passive mode and behaves as the initial condition of the CMB fluctuations after reentering the horizon \citep{Shaw:2009nf}.

Assuming Gaussianity of PMFs, the induced GW, $h_{ij} \propto (B_i B_j)^{TT}$, becomes a chi-square random field \citep{Brown:2005kr}. When the PMF power spectrum has a scale-invariant shape, in analogy with the usual local type NG, the tensor-tensor-tensor bispectrum has the squeezed shape \citep{Shiraishi:2011dh,Shiraishi:2012rm}: 
\begin{equation} 
  \mathcal{B}_{k_1 k_2 k_3}^{\lambda_1 \lambda_2 \lambda_3}
  = \sqrt{2} f_{\rm NL}^{ttt, {\rm sq}} S_{k_1 k_2 k_3}^{\rm loc} ,  \label{eq:B_PMF}
\end{equation}
where the usual scalar local bispectrum template reads:
\begin{equation}
  S_{k_1 k_2 k_3}^{\rm loc}
  \equiv \frac{6}{5} \left( 2 \pi^2 \mathcal{P}_\zeta \right)^2
  \left(\frac{1}{k_1^{3} k_2^{3}} + 2~{\rm perm} \right) , 
\end{equation}
and $A_{\hat{k}_1 \hat{k}_2 \hat{k}_3}^{\lambda_1 \lambda_2 \lambda_3}$ is given in the same form as Eq.~\eqref{eq:A_pseudo}. A newly introduced parameter $f_{\rm NL}^{ttt, \rm sq}$ represents the relative size of the $\lambda = +2$ bispectrum to $S_{k_1 k_2 k_3}^{\rm loc}$ and satisfies: 
\begin{equation}
  f_{\rm NL}^{ttt, \rm sq} \equiv \lim_{ \substack{k_1 \to 0 \\ k_2 \to k_3}}
  \frac{\mathcal{B}_{k_1 k_2 k_3}^{+2+2+2} A_{\hat{k}_1 \hat{k}_2 \hat{k}_3}^{+2 +2 +2} }
       {S_{k_1 k_2 k_3}^{\rm loc} } . \label{eq:fnl_def_ttt_sq} 
\end{equation}
As the GW bispectrum has a sextic dependence on PMFs, $f_{\rm NL}^{ttt, \rm sq}$ takes the form:
\begin{equation}
  f_{\rm NL}^{ttt, \rm sq} \sim \left( \frac{B_{1 \, \rm Mpc}}{1 \, \rm nG} \right)^6, \label{eq:fnl_ttt_sq_PMF}
\end{equation}
where $B_{1 \, \rm Mpc}$ is the PMF strength smoothed on $1 \, {\rm Mpc}$. This means that $f_{\rm NL}^{ttt, \rm sq} > 1$ is realized if nano-Gauss-level PMFs exist.
Note that, in the same mechanism, the scalar local type bispectrum is generated, however, it is subdominant compared with the GW one \citep{Shiraishi:2013vha}.

This is the result under the absence of a helical (parity-odd) term in the PMF power spectrum, however, the bispectrum shape is modified if it exists \citep{Shiraishi:2012sn}.

\subsection{Modified gravity}

The above scenarios rely on some extra source fields. Let us discuss another NG GW production through the modification of Einstein gravity.

\subsubsection{Weyl gravity} \label{subsubsec:Weyl}

In the context of the Weyl gravity, there may exist the Chern-Simons term $f(\phi) \widetilde{W}W$, where $W$ is the Weyl tensor and $\widetilde{W}$ is its dual. Like the axion inflation models, this term also sources chiral GWs \citep{Lue:1998mq,Alexander:2004wk}. The tensor-tensor-scalar bispectrum, whose angular structure reads:
\begin{equation}
  A_{\hat{k}_1 \hat{k}_2 \hat{k}_3}^{\lambda_1 \lambda_2 0}
  = 
  (\hat{k}_1 \cdot \hat{k}_2) \,
  e_{ij}^{(-\lambda_1)}(\hat{k}_1) e_{ij}^{(-\lambda_2)}(\hat{k}_2) ,
  \label{eq:A_WW}
\end{equation}
surpasses the tensor-tensor-tensor and tensor-scalar-scalar ones in amplitude \citep{Bartolo:2017szm}. Due to the parity-violating nature in the GW sector, the helical bispectrum obeys
\begin{equation}
  \mathcal{B}_{k_1 k_2 k_3}^{\lambda_1 \lambda_2 0} \propto \frac{\lambda_1}{2}  \delta_{\lambda_1, \lambda_2} . \label{eq:B_WW}
\end{equation}

One can also construct the cubic actions $f(\phi) W^3$ and $f(\phi) \widetilde{W}W^2$ \citep{Maldacena:2011nz,Soda:2011am,Shiraishi:2011st}. The tensor-tensor-tensor bispectra generated from both terms have identical angular dependence as Eq.~\eqref{eq:A_pseudo}. In contrast, the helicity dependence is completely different \citep{Shiraishi:2011st}: the former gives rise to the non-helical (parity-even) contribution:
\begin{equation}
\mathcal{B}_{k_1 k_2 k_3}^{\lambda_1 \lambda_2 \lambda_3} \propto \delta_{\lambda_1, \lambda_2} \delta_{\lambda_2, \lambda_3}, \label{eq:B_W3}
\end{equation}
while the latter induces the helical (parity-odd) one
\begin{equation}
\mathcal{B}_{k_1 k_2 k_3}^{\lambda_1 \lambda_2 \lambda_3} \propto \frac{\lambda_1}{2}
\delta_{\lambda_1, \lambda_2} \delta_{\lambda_2, \lambda_3} . \label{eq:B_WW2}
\end{equation}

\subsubsection{Massive gravity} \label{subsubsec:massive}

In single-field slow-roll inflation with Einstein gravity, the tensor-scalar-scalar bispectrum is affected by slow-roll suppression \citep{Maldacena:2002vr}. However, in a massive gravity model \citep{Domenech:2017kno}, the size of the bispectrum is controlled by the graviton mass. The bispectrum, taking the form:
\begin{eqnarray}
  \mathcal{B}_{k_1 k_2 k_3}^{\lambda_1 ~ 0 ~ 0 }
  &=& - \frac{8 \sqrt{2}}{5}  f_{\rm NL}^{tss, \rm sq}
\left(2 \pi^2 \mathcal{P}_\zeta \right)^2
 \frac{1}{k_1^3 k_2^2 k_3^2}
\left[- k_t + \frac{k_1 k_2 + k_2 k_3 + k_3
  k_1}{k_t} + \frac{k_1 k_2 k_3}{k_t^2} \right] , \label{eq:B_massive}  \\
  A_{\hat{k}_1 \hat{k}_2 \hat{k}_3}^{\lambda_1 ~ 0 ~ 0 }
  &=& e_{ij}^{(-\lambda_1)}(\hat{k}_1) \hat{k}_{2i} \hat{k}_{3j} , \label{eq:A_massive}
\end{eqnarray}
with $k_t \equiv k_1 + k_2 + k_3$, is amplified at long-wavelength tensor and short-wavelength scalar configurations ($k_1 \ll k_2 \sim k_3$). A newly introduced parameter $f_{\rm NL}^{tss, \rm sq}$ denotes the relative size of the $\lambda = +2$ bispectrum to $S_{k_1 k_2 k_3}^{\rm loc}$, obeying: 
\begin{equation}
  f_{\rm NL}^{tss, \rm sq} \equiv
  \lim_{ \substack{k_1 \to 0 \\ k_2 \to k_3}}
  \frac{\mathcal{B}_{k_1 k_2 k_3}^{+2 ~ 0 ~ 0 } A_{\hat{k}_1 \hat{k}_2 \hat{k}_3}^{+2 ~ 0 ~ 0 }}
       {S_{k_1 k_2 k_3}^{\rm loc}}. \label{eq:fnl_def_tss_sq} 
\end{equation}
With an upper bound on the tensor-to-scalar ratio $r \lesssim 0.1$, the prediction in Einstein gravity, $f_{\rm NL}^{tss, \rm sq} \sim 0.1 \, r$, indicates $f_{\rm NL}^{tss, \rm sq} \lesssim 0.01$. In contrast, massive gravity model modifies this as:
\begin{equation}
  f_{\rm NL}^{tss, \rm sq} \sim 0.1 \, r \lambda_{sst} , \label{eq:fnl_tss_sq_massive}
\end{equation}
and $f_{\rm NL}^{tss, \rm sq} > 1$ can then be realized depending on the strength parameter of a non-linear interaction $\lambda_{sst}$. Note that the spectral index of the GW power spectrum is also a possible observable of this model \citep{Domenech:2017kno}, but has been unconstrained so far.

\section{CMB bispectra from tensor non-Gaussianities} \label{sec:CMB_bis}

Next, let us discuss the CMB bispectra generated from GW NGs.


\subsection{General formalism for CMB angular bispectrum}

We start by reviewing how to compute the CMB bispectra generated from the primordial scalar, vector, and tensor NGs based on the general formalism developed in Refs.~\citep{Shiraishi:2010sm,Shiraishi:2010kd}.

The CMB field is characterized by temperature and two linear polarizations called $E$ and $B$ modes. All of these are spin-0 fields, and the temperature and $E$-mode fields have even parity, while the $B$ mode is parity-odd. Since temperature ($X = T$), $E$-mode ($X = E$), and $B$-mode ($X = B$) fields are distributed on the 2D sphere, they can be expanded using the spherical harmonic basis according to
\begin{equation}
  X(\hat{n}) = \sum_{\ell m} a_{\ell m}^X Y_{\ell m}(\hat{n}) ,
\end{equation}
where we have assumed the spatial flatness of the Universe. The spherical harmonic coefficients $a_{\ell m}^X$ are given by the sum of the scalar, vector, and tensor modes as:  
\begin{equation}
  a_{\ell m}^X  = a_{\ell m}^{X (s)} + a_{\ell m}^{X (v)} + a_{\ell m}^{X (t)}.
\end{equation}
Note that the scalar mode cannot generate a $B$-mode signal due to zero helicity. The coefficients of the scalar ($z = s$), vector ($z = v$), and tensor ($z = t$) modes have the structure:
\begin{equation}
a^{X (z)}_{\ell m} = 4\pi (-i)^\ell 
\int \frac{d^3{\bf k}}{(2\pi)^3} \mathcal{T}_{ \ell, k}^{X (z)}
\sum_{\lambda} [{\rm sgn}(\lambda)]^{\lambda+x}  \xi_{\bf k}^{(\lambda)} 
{}_{-\lambda}Y_{\ell m}^*(\hat{k}) ~, \label{eq:alm_general}
\end{equation}
where $x$ changes depending on the parity of the CMB field according to $x = 0$ for $X = T,E$ and $x = 1$ for $X = B$, and recall that $\lambda$ represents helicity as $\lambda = 0$ for $z = s$, $\lambda = \pm 1$ for $z = v$, and $\lambda = \pm 2$ for $z = t$. The linear transfer function $\mathcal{T}_{\ell, k}^{X (z) }$ represents the time evolution of the CMB fluctuations originating from the primordial perturbation $\xi_{\bf k}^{(\lambda)}$. Using Eq.~\eqref{eq:alm_general}, one can form the CMB bispectrum as:
\begin{equation}
  \Braket{\prod_{n = 1}^{3} a^{X_n (z_n)}_{\ell_n m_n} }
  = \left[\prod_{n = 1}^{3} (-i)^{\ell_n} 
\int \frac{d^3{\bf k}_n}{2\pi^2} \mathcal{T}_{\ell_n, k_n}^{X_n (z_n)}
\sum_{\lambda_n} [{\rm sgn}(\lambda_n)]^{\lambda_n + x_n} {}_{-\lambda_n}Y_{\ell_n m_n}^*(\hat{k}_n) \right] 
  \Braket{\prod_{n = 1}^{3} \xi_{{\bf k}_n}^{(\lambda_n)} }. 
\end{equation}
The primordial curvature perturbation and the primordial GW act as initial conditions of the scalar and tensor CMB anisotropies, respectively; thus, $\xi_{{\bf k}}^{(0)} = \zeta_{{\bf k}}$ and $\xi_{{\bf k}}^{(\pm 2)} = h_{{\bf k}}^{(\pm 2)}$. Employing the harmonic expansion:
\begin{equation}
  \xi_{{\bf k}}^{(\lambda)}
  = \sum_{\ell m} \xi_{\ell m}^{(\lambda)}(k) {}_{-\lambda}Y_{\ell m}(\hat{k}) ,
\end{equation}
the above formula is rewritten as:
\begin{equation}
   \Braket{\prod_{n = 1}^{3} a^{X_n (z_n)}_{\ell_n m_n} }
  = \left[\prod_{n = 1}^{3} (-i)^{\ell_n} 
\int_0^{\infty} \frac{k_n^2 dk_n}{2\pi^2} \mathcal{T}_{\ell_n, k_n}^{X_n (z_n)}
\sum_{\lambda_n} [{\rm sgn}(\lambda_n)]^{\lambda_n + x_n}  \right] 
  \Braket{\prod_{n = 1}^{3}  \xi_{\ell_n m_n}^{(\lambda_n)}(k_n) }, \label{eq:alm3_general} 
\end{equation}
where,
\begin{equation}
  \Braket{\prod_{n = 1}^{3}  \xi_{\ell_n m_n}^{(\lambda_n)}(k_n) }
  = \left[\prod_{n = 1}^{3} \int d^2 \hat{k}_n \, {}_{-\lambda_n}Y_{\ell_n m_n}^*(\hat{k}_n) \right] \Braket{\prod_{n = 1}^{3} \xi_{{\bf k}_n}^{(\lambda_n)} }.
  \label{eq:xi3}
\end{equation}
This is further simplified once the explicit formula of $\Braket{\prod_{n = 1}^{3} \xi_{{\bf k}_n}^{(\lambda_n)} }$ is given. The computation procedure is as follows:
\begin{enumerate}
\item Expand $\Braket{\prod_{n = 1}^{3} \xi_{{\bf k}_n}^{(\lambda_n)} }$ using the (spin-weighted) spherical harmonics in terms of $\hat{k}_1$, $\hat{k}_2$, and $\hat{k}_3$.
\item Perform the angular integrals, $\int d^2 \hat{k}_1$, $\int d^2 \hat{k}_2$, and $\int d^2 \hat{k}_3 $, of the resultant spherical harmonics and convert them into products of Wigner symbols.
\item Simplify the resultant products of the Wigner symbols by adding angular momenta.
\end{enumerate}

If the primordial bispectrum $\Braket{\prod_{n = 1}^{3} \xi_{{\bf k}_n}^{(\lambda_n)} }$ respects rotational invariance, via the above computation, the Wigner $3j$ symbol
$\left(
\begin{array}{ccc}
  \ell_1 & \ell_2 & \ell_3 \\
   m_1 & m_2 & m_3
   \end{array}
\right)$
is singled out in $\Braket{\prod_{n = 1}^{3}  \xi_{\ell_n m_n}^{(\lambda_n)}(k_n)}$ and hence the resulting CMB bispectrum takes the form:
\begin{equation}
  \Braket{\prod_{n = 1}^{3} a^{X_n (z_n)}_{\ell_n m_n} }
  = \left(
  \begin{array}{ccc}
  \ell_1 & \ell_2 & \ell_3 \\
   m_1 & m_2 & m_3
  \end{array}
\right) B_{(z_1 z_2 z_3) \ell_1 \ell_2 \ell_3}^{X_1 X_2 X_3}. \label{eq:alm3_rot_inv}
\end{equation}
In this case, the nonzero signal is confined to the tetrahedral domain:
\begin{equation}
  |\ell_1 - \ell_2| \leq \ell_3 \leq \ell_1 + \ell_2 . \label{eq:L_tetra}
\end{equation}


\subsection{Practical examples}

In the following, we demonstrate the CMB bispectrum computation in the context of the practical cases discussed in Sec.~\ref{sec:models}.

\subsubsection{Three tensors}

First, we discuss the CMB bispectra sourced by primordial tensors where we examine the axion model $f_{\rm NL}^{ttt, \rm eq}$ template~\eqref{eq:B_pseudo} \citep{Shiraishi:2013kxa}, the $W^3$ model~\eqref{eq:B_W3} \citep{Shiraishi:2011st}, the $\widetilde{W}W^2$ model~\eqref{eq:B_WW2} \citep{Shiraishi:2011st}, and the PMF model $f_{\rm NL}^{ttt, \rm sq}$~\eqref{eq:B_PMF} \citep{Shiraishi:2011dh}. Since $A_{\hat{k}_1 \hat{k}_2 \hat{k}_3}^{\lambda_1 \lambda_2 \lambda_3}$ takes the identical form, $\Braket{\prod_{n = 1}^{3}  h_{\ell_n m_n}^{(\lambda_n)}(k_n) }$ can be computed in a similar way.

For the first step, in order to simplify the R.H.S. of Eq.~\eqref{eq:xi3}, we expand $A_{\hat{k}_1 \hat{k}_2 \hat{k}_3}^{\lambda_1 \lambda_2 \lambda_3}$ and the Dirac delta function:    
\begin{eqnarray}
  e_{ij}^{(-\lambda_1)}(\hat{k}_1) e_{jk}^{(-\lambda_2)}(\hat{k}_2) e_{ki}^{(-\lambda_3)}(\hat{k}_3)
&=& -  \frac{( 8\pi )^{3/2}}{10} \sqrt{\frac{7}{3}}
\left[ \prod_{n = 1}^3 \sum_{\mu_n} {}_{\lambda_n}Y_{2 \mu_n}^*(\hat{k}_n) \right]
\left(
  \begin{array}{ccc}
   2 & 2 & 2 \\
  \mu_1 & \mu_2 & \mu_3
  \end{array}
  \right) , \\
\delta^{(3)} \left( \sum_{n=1}^3 {\bf k}_n \right) 
&=& 8 \int_0^\infty y^2 dy 
\left[ \prod_{n=1}^3 \sum_{L_n M_n} 
 (-1)^{L_n/2} j_{L_n}(k_n y) 
Y_{L_n M_n}^*(\hat{k}_n) \right] \nonumber \\ 
&& \times h_{L_1 L_2 L_3}^{0 \ 0 \ 0}
 \left(
  \begin{array}{ccc}
  L_1 & L_2 & L_3 \\
  M_1 & M_2 & M_3 
  \end{array}
 \right) ,
\end{eqnarray}
where 
\begin{equation}
h^{s_1 s_2 s_3}_{l_1 l_2 l_3} 
\equiv \sqrt{\frac{(2 l_1 + 1)(2 l_2 + 1)(2 l_3 + 1)}{4 \pi}}
\left(
  \begin{array}{ccc}
  l_1 & l_2 & l_3 \\
   s_1 & s_2 & s_3 
  \end{array}
  \right).
\end{equation}
For the second step, the angular integrals of the resultant spherical harmonics are performed, such that:
\begin{equation}
  \int d^2 \hat{k}_n \,
  {}_{-\lambda_n}Y_{\ell_n m_n}^*(\hat{k}_n) \,
  Y_{L_n M_n}^*(\hat{k}_n) \,
  {}_{\lambda_n}Y_{2 \mu_n}^*(\hat{k}_n) 
  = h_{\ell_n L_n 2}^{\lambda_n 0 -\lambda_n}
  \left(
  \begin{array}{ccc}
    \ell_n & L_n & 2 \\
    m_n & M_n & \mu_n
  \end{array}
  \right). 
\end{equation}
As a final step, the summation of the Wigner $3j$ symbols appearing above over angular momenta is computed according to:
\begin{eqnarray}
&&  \sum_{\substack{M_1 M_2 M_3 \\ \mu_1 \mu_2 \mu_3}} \left(
  \begin{array}{ccc}
  L_1 & L_2 & L_3 \\
  M_1 & M_2 & M_3 
  \end{array}
  \right)
  \left(
  \begin{array}{ccc}
   2 & 2 & 2 \\
  \mu_1 & \mu_2 & \mu_3
  \end{array}
  \right)
  \left(
  \begin{array}{ccc}
   \ell_1 & L_1 & 2 \\
  m_1 & M_1 & \mu_1
  \end{array}
  \right)
  \nonumber \\
  && \times
  \left(
  \begin{array}{ccc}
   \ell_2 & L_2 & 2 \\
  m_2 & M_2 & \mu_2
  \end{array}
  \right)
  \left(
  \begin{array}{ccc}
   \ell_3 & L_3 & 2 \\
  m_3 & M_3 & \mu_3
  \end{array}
 \right)
 =  \left(
  \begin{array}{ccc}
   \ell_1 & \ell_2 & \ell_3 \\
  m_1 & m_2 & m_3
  \end{array}
  \right)
  \left\{
  \begin{array}{ccc}
    \ell_1 & \ell_2 & \ell_3 \\
    L_1 & L_2 & L_3 \\
    2 & 2 & 2
  \end{array}
  \right\}.
\end{eqnarray}
Consequently, we obtain:
\begin{eqnarray}
  \Braket{\prod_{n=1}^3 h_{\ell_n m_n}^{(\lambda_n)}(k_n) }
&=&  \left(
  \begin{array}{ccc}
   \ell_1 & \ell_2 & \ell_3 \\
  m_1 & m_2 & m_3
  \end{array}
  \right) \times
  \frac{- ( 8\pi )^{3/2}}{10} \sqrt{\frac{7}{3}}
  \sum_{L_1 L_2 L_3} (-1)^{\frac{L_1 + L_2 + L_3}{2}}  h_{L_1 L_2 L_3}^{0 \ 0 \ 0}
  \nonumber \\
   &&   
    \times \left\{
  \begin{array}{ccc}
    \ell_1 & \ell_2 & \ell_3 \\
    L_1 & L_2 & L_3 \\
    2 & 2 & 2
  \end{array}
  \right\} 
  \int_0^\infty y^2 dy
  \left[ \prod_{n = 1}^3 4\pi  
    j_{L_n}(k_n y) h_{\ell_n L_n 2}^{\lambda_n 0 -\lambda_n} \right]
   \mathcal{B}_{k_1 k_2 k_3}^{\lambda_1 \lambda_2 \lambda_3} .
\end{eqnarray}
Here, the prefactor $\left(
  \begin{array}{ccc}
   \ell_1 & \ell_2 & \ell_3 \\
  m_1 & m_2 & m_3
  \end{array}
  \right)$
  is due to rotational invariance of the primordial bispectrum. Inserting this into Eq.~\eqref{eq:alm3_general} and simplifying the resultant equation, we find that the CMB bispectra take the rotational-invariant form \eqref{eq:alm3_rot_inv} with:
  \begin{eqnarray}
    B_{(ttt) \ell_1 \ell_2 \ell_3}^{X_1 X_2 X_3}
    &=&
    \frac{- ( 8\pi )^{3/2}}{10} \sqrt{\frac{7}{3}} (-i)^{\ell_1 + \ell_2 + \ell_3} 
    \sum_{L_1 L_2 L_3} (-1)^{\frac{L_1 + L_2 + L_3}{2}}  h_{L_1 L_2 L_3}^{0 \ 0 \ 0}
    \left\{
    \begin{array}{ccc}
      \ell_1 & \ell_2 & \ell_3 \\
      L_1 & L_2 & L_3 \\
    2 & 2 & 2
    \end{array}
    \right\} \nonumber \\
    && \times  \int_0^\infty y^2 dy
    \left[\prod_{n = 1}^{3}  \frac{2}{\pi} 
      \int_0^{\infty} k_n^2 dk_n \mathcal{T}_{ \ell_n, k_n}^{X_n (t)} 
      j_{L_n}(k_n y) \right] \nonumber \\
    &&
    \times \left[\prod_{n = 1}^{3} \sum_{\lambda_n = \pm 2} [{\rm sgn}(\lambda_n)]^{x_n} 
      h_{\ell_n L_n 2}^{\lambda_n 0 -\lambda_n} \right]
    \mathcal{B}_{k_1 k_2 k_3}^{\lambda_1 \lambda_2 \lambda_3} .
  \end{eqnarray}

The $\ell$-space domain containing the non-zero signal is determined by the summation over the helicities $\lambda_1$, $\lambda_2$, and $\lambda_3$. In the $\widetilde{W}W^2$ model, performing the summation of Eq.~\eqref{eq:B_WW2} leads to the non-vanishing condition: \citep{Shiraishi:2011st}
  \begin{equation}
    \ell_1 + \ell_2 + \ell_3 + x_1 + x_2 + x_3 = {\rm odd} .
  \end{equation}
  On the other hand, in the $W^3$ and PMF models, due to the helicity dependence in Eqs.~\eqref{eq:B_W3} and \eqref{eq:B_PMF}, the non-vanishing signal is confined to \citep{Shiraishi:2011dh,Shiraishi:2011st}:
    \begin{equation}
    \ell_1 + \ell_2 + \ell_3 + x_1 + x_2 + x_3 = {\rm even} .
    \end{equation}    
  In the axion model, however, the helicity dependence in Eq.~\eqref{eq:B_pseudo} does not yield any restriction, allowing a non-vanishing signal in the whole tetrahedral domain \citep{Shiraishi:2013kxa}.

  The left and center panels of Fig.~\ref{fig:3D} show the intensity distributions of the temperature bispectra from the axion model $f_{\rm NL}^{ttt, \rm eq}$ template \eqref{eq:B_pseudo} and the PMF model $f_{\rm NL}^{ttt, \rm sq}$ template \eqref{eq:B_PMF}, respectively. As expected, it is confirmed that the signal comes mostly from the equilateral ($\ell_1 \sim \ell_2 \sim \ell_3$) and squeezed ($\ell_1 \ll \ell_2 \sim \ell_3$, $\ell_2 \ll \ell_3 \sim \ell_1$ and $\ell_3 \ll \ell_1 \sim \ell_2$) configurations, respectively. The decaying nature for $\ell \gtrsim 100$ due to the lack of the tensor-mode integrated Sachs-Wolfe contribution \citep{Shiraishi:2011dh,Shiraishi:2013kxa} is also visually apparent.

\subsubsection{Two tensors and one scalar}

Here, we compute the CMB tensor-tensor-scalar bispectrum from the $\widetilde{W}W$ model \eqref{eq:B_WW} \citep{Bartolo:2018elp}. The angular dependence in $A_{\hat{k}_1 \hat{k}_2 \hat{k}_3}^{\lambda_1 \lambda_2 0}$ \eqref{eq:A_WW} is decomposed according to: 
  \begin{equation}
 (\hat{k}_1 \cdot \hat{k}_2) e_{ij}^{(-\lambda_1)}(\hat{k}_1) e_{ij}^{(-\lambda_2)}(\hat{k}_2)
    = \frac{32 \pi^2}{15}
    \sum_{J \mu} {}_{\lambda_1}Y_{J \mu}^*(\hat{k}_1)  
        {}_{\lambda_2}Y_{J -\mu}^*(\hat{k}_2) 
        (-1)^{\mu + 1 + J}
        \frac{h_{1 ~ 2 ~J}^{0 \lambda_1 -\lambda_1} h_{1 ~ 2 ~ J}^{0 \lambda_2 -\lambda_2}}{2J + 1} .
  \end{equation}
  By means of the above methodology, we obtain \citep{Bartolo:2018elp}:
\begin{eqnarray}
  B_{(tts) \ell_1 \ell_2 \ell_3}^{X_1 X_2 X_3}
  &=& \frac{32 \pi^2}{15}
(-i)^{\ell_1 + \ell_2 + \ell_3} \sum_{L_1 L_2} (-1)^{\frac{L_1 + L_2 + \ell_3}{2}}
  h_{L_1 L_2 \ell_3}^{0~0~0} 
  \sum_{J} \frac{ (-1)^{1 + J + L_2 +  \ell_1 } }{2J + 1}
       \left\{
  \begin{array}{ccc}
  \ell_1 & \ell_2 & \ell_3 \\
  L_2 & L_1 & J 
  \end{array}
  \right\}
\nonumber \\
&& \times \int_0^\infty y^2 dy
\left[ \prod_{n=1}^2 \frac{2}{\pi}  \int_0^\infty k_n^2 dk_n 
  \mathcal{T}_{\ell_n, k_n }^{X_n (t)} j_{L_n}(k_n y)   \right]
\frac{2}{\pi}  \int_0^\infty k_3^2 d k_3
 \mathcal{T}_{\ell_3, k_3}^{X_3 (s)} j_{\ell_3}(k_3 y) 
 \nonumber \\
 && \times \left[\prod_{n=1}^2 \sum_{\lambda_n = \pm 2} \left[{\rm sgn}(\lambda_n)\right]^{x_n}  h_{L_n ~ \ell_n ~ J}^{0 \lambda_n -\lambda_n}  h_{1 ~ 2 ~J}^{0 \lambda_n -\lambda_n} \right]
 \mathcal{B}_{k_1 k_2 k_3}^{\lambda_1 \lambda_2 0} .
\end{eqnarray}
Performing the summation over $\lambda_1$ and $\lambda_2$ with Eq.~\eqref{eq:B_WW}, we find that the non-vanishing signal obeys \citep{Bartolo:2018elp}:
\begin{equation}
  \ell_1 + \ell_2 + \ell_3 + x_1 +x_2 = {\rm odd}.
\end{equation}

\subsubsection{One tensor and two scalars}

We here compute the CMB bispectrum from the massive gravity model $f_{\rm NL}^{tss, \rm sq}$ template \eqref{eq:B_massive} \citep{Domenech:2017kno}. The spherical harmonic expansion of $A_{\hat{k}_1 \hat{k}_2 \hat{k}_3}^{\lambda_1 ~0 ~0}$ \eqref{eq:A_massive} then reads: 
\begin{equation}
  e_{ij}^{(-\lambda_1)}(\hat{k}_1) \hat{k}_{2i} \hat{k}_{3j}
  = \frac{(8\pi)^{3/2}}{6}     
  \sum_{\mu_1 \mu_2 \mu_3} {}_{\lambda_1}Y_{2 \mu_1}^*(\hat{k}_1) \,
  Y_{1 \mu_2}^*(\hat{k}_2) \,
  Y_{1 \mu_3}^*(\hat{k}_3)
\left(
  \begin{array}{ccc}
  2 & 1 &  1\\
  \mu_1 & \mu_2 & \mu_3 
  \end{array}
  \right) .
\end{equation}
In the same manner, we can derive the CMB bispectrum as \citep{Shiraishi:2010kd,Domenech:2017kno}:
\begin{eqnarray}
   B_{(tss) \ell_1 \ell_2 \ell_3}^{X_1 X_2 X_3}
   &=&  \frac{(8\pi)^{3/2}}{6} 
   (-i)^{\ell_1 + \ell_2 + \ell_3}
   \sum_{L_1 L_2 L_3}  (-1)^{\frac{L_1 + L_2 + L_3}{2}} 
   h_{L_1 L_2 L_3}^{0 \ 0 \ 0}
   h_{\ell_2 L_2 1}^{0~ 0 ~0}
   h_{\ell_3 L_3 1}^{0~ 0 ~0}
   \left\{
  \begin{array}{ccc}
    \ell_1 & \ell_2 & \ell_3 \\
    L_1 & L_2 & L_3 \\
    2 & 1 &  1
  \end{array}
  \right\}
   \nonumber \\  
   && \times
   \int_0^\infty y^2 dy 
   \frac{2}{\pi} \int_0^{\infty}  k_1^2 dk_1 \mathcal{T}_{\ell_1, k_1}^{X_1 (t)}
   j_{L_1}(k_1 y) 
   \left[\prod_{n = 2}^{3}
     \frac{2}{\pi} \int_0^{\infty}  k_n^2 dk_n \mathcal{T}_{\ell_n, k_n}^{X_n (s)}
     j_{L_n}(k_n y) 
  \right]
   \nonumber \\ 
&& \times
\sum_{\lambda_1 = \pm 2} [{\rm sgn}(\lambda_1)]^{x_1} h_{\ell_1 L_1 2}^{\lambda_1 0 -\lambda_1}  \mathcal{B}_{k_1 k_2 k_3}^{\lambda_1 0 ~ 0 } . 
\end{eqnarray}
By summing over $\lambda_1$, we confirm that, because of the absence of the helicity dependence in Eq.~\eqref{eq:B_massive}, the non-vanishing signal is distributed on \citep{Shiraishi:2010kd,Domenech:2017kno}:
\begin{equation}
  \ell_1 + \ell_2 + \ell_3 + x_1 = {\rm even}.
\end{equation}

\begin{figure*}[t]
  \begin{tabular}{ccc} 
   \begin{minipage}{0.33\hsize}
     \begin{center}
       \includegraphics[width=1\textwidth]{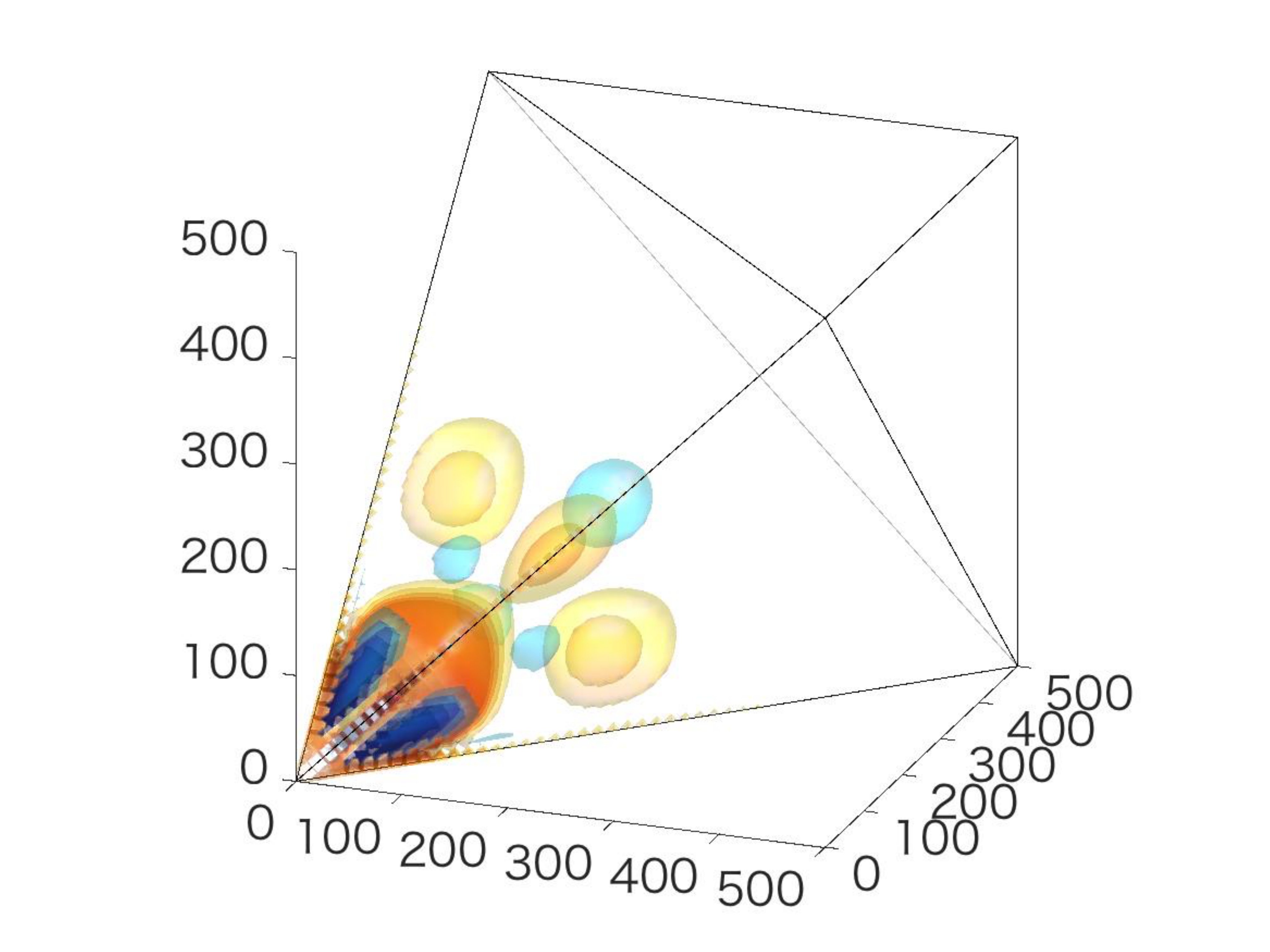}
       $f_{\rm NL}^{ttt, \rm eq}$
  \end{center}
   \end{minipage}
   \begin{minipage}{0.33\hsize}
     \begin{center}
       \includegraphics[width=1\textwidth]{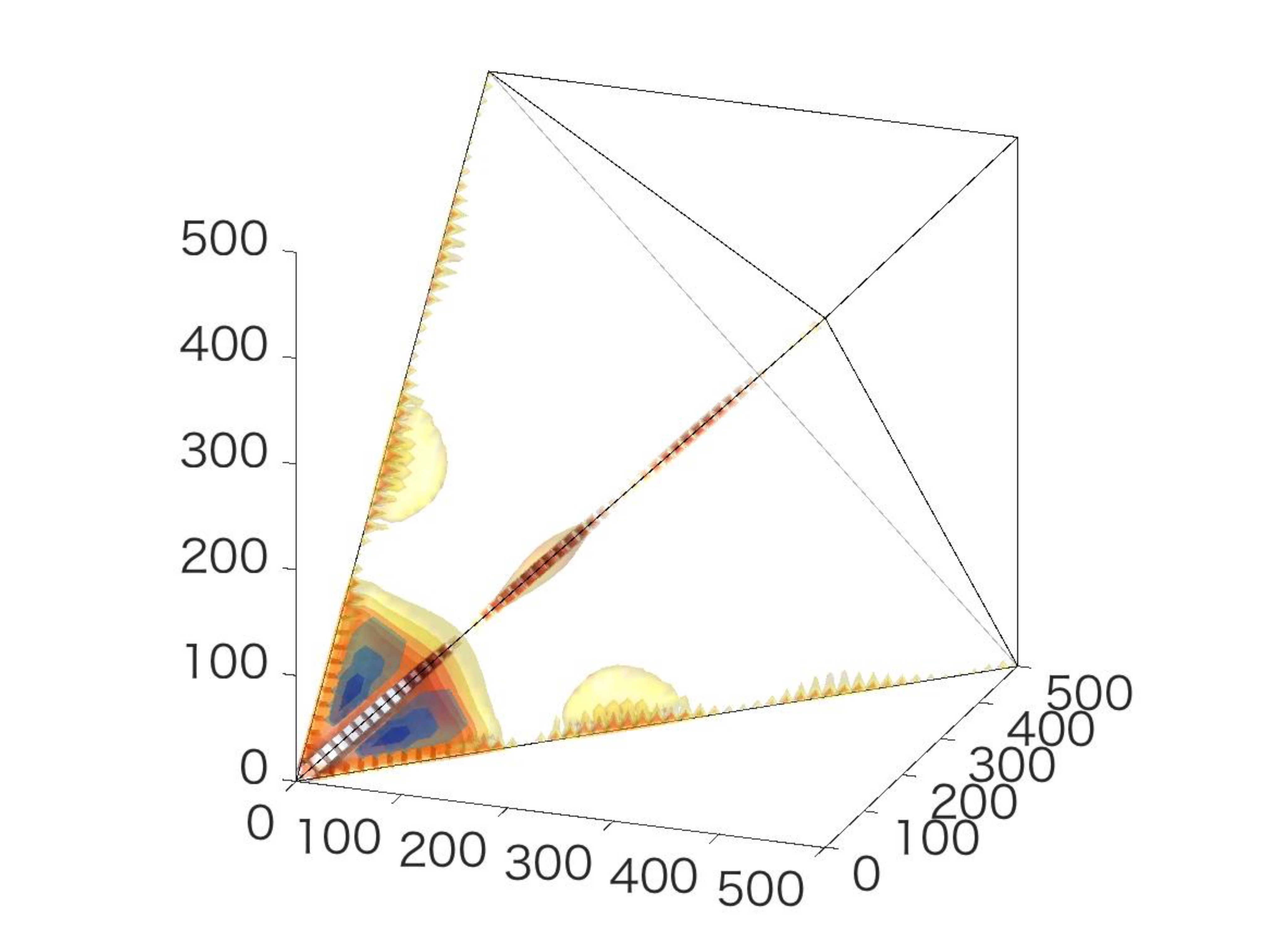}
       $f_{\rm NL}^{ttt, \rm sq}$
  \end{center}
   \end{minipage}
   \begin{minipage}{0.33\hsize}
     \begin{center}
       \includegraphics[width=1\textwidth]{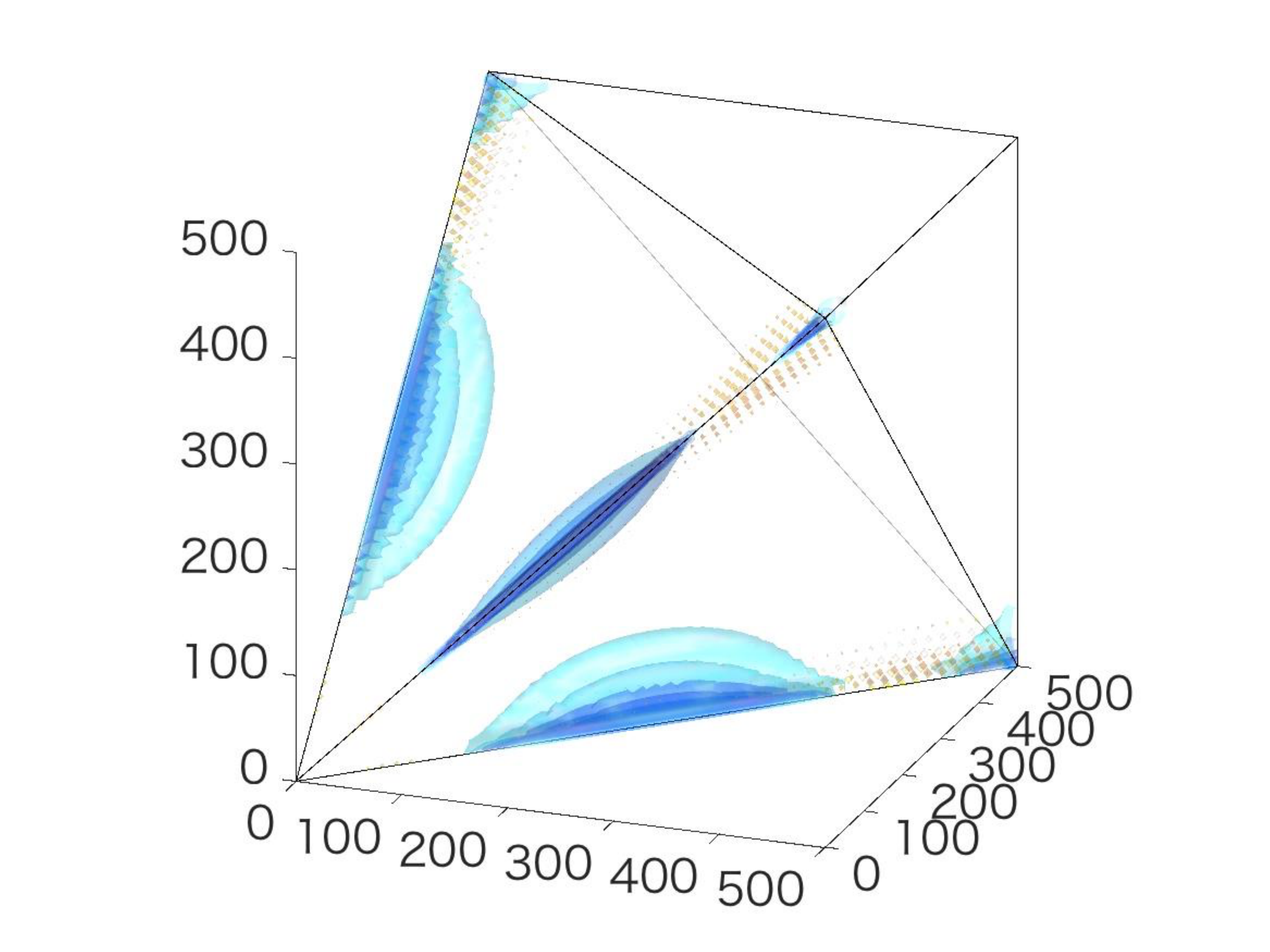}
       $f_{\rm NL}^{tss, \rm sq}$
     \end{center}
   \end{minipage}
  \end{tabular}
  \caption{Intensity distributions of the CMB temperature bispectra from the axion model $f_{\rm NL}^{ttt, \rm eq}$ template \eqref{eq:B_pseudo}, the PMF model $f_{\rm NL}^{ttt, \rm sq}$ one \eqref{eq:B_PMF}, and the massive gravity model $f_{\rm NL}^{tss, \rm sq}$ one \eqref{eq:B_massive} in the $\ell$-space tetrahedral domain where the axes correspond to $\ell_1$, $\ell_2$, and $\ell_3$, respectively. Here, the bispectra are rescaled using a constant Sachs-Wolfe template \citep{Fergusson:2009nv}. Dense red (blue) color expresses a larger positive (negative) signal. The current and future observational limits on $f_{\rm NL}^{ttt, \rm eq}$, $f_{\rm NL}^{ttt, \rm sq}$, and $f_{\rm NL}^{tss, \rm sq}$ are discussed in Secs.~\ref{subsec:limits} and \ref{subsec:future}, respectively.}
\label{fig:3D}
\end{figure*}

The right panel of Fig.~\ref{fig:3D} describes the intensity distribution of the resultant temperature bispectrum showing explicitly that it is amplified in the squeezed limit. Differently from the $f_{\rm NL}^{ttt, \rm sq}$ case, the squeezed signal survives even for $\ell \gtrsim 100$ because of the presence of the scalar-mode acoustic oscillation.

\section{Measurements of CMB tensor-mode bispectra} \label{sec:data}

In this section, we examine the present and future constraints on GW NGs through the CMB bispectrum measurements.

\subsection{Optimal tensor-mode bispectrum estimation}

Let us begin by explaining how to estimate the overall magnitude of the primordial bispectrum, dubbed as $f_{\rm NL}$, from the CMB data.

An optimal $f_{\rm NL}$ estimator takes the form \citep{Komatsu:2003iq}:
\begin{equation}
  \hat{f}_{\rm NL} = \frac{1}{F}
  \sum_{\ell_1 \ell_2 \ell_3} 
 (-1)^{\ell_1 + \ell_2 + \ell_3}  
 \frac{B_{\ell_1 \ell_2 \ell_3}^{\rm th} B_{\ell_1 \ell_2 \ell_3}^{\rm obs}}{6 C_{\ell_1} C_{\ell_2} C_{\ell_3}} ,  \label{eq:estimator_general}
\end{equation}
where $C_\ell$ is the CMB angular power spectrum, $B_{\ell_1 \ell_2 \ell_3}^{\rm th}$ is the theoretical template of the angle-averaged CMB bispectrum one wants to measure [corresponding to $B_{(z_1 z_2 z_3) \ell_1 \ell_2 \ell_3}^{X_1 X_2 X_3}$ in Eq.~\eqref{eq:alm3_rot_inv} for $f_{\rm NL} = 1$],
\begin{equation}
  B_{\ell_1 \ell_2 \ell_3}^{\rm obs}
  \equiv \sum_{m_1 m_2 m_3} \left(
\begin{array}{ccc}
\ell_1 & \ell_2 & \ell_3 \\
m_1 & m_2 & m_3
\end{array}
\right)
\left[
a_{\ell_1 m_1} a_{\ell_2 m_2} a_{\ell_3 m_3}
- \left(\Braket{a_{\ell_1 m_1} a_{\ell_2 m_2}}a_{\ell_3 m_3} + 2 \ {\rm perm} \right)
\right]
\end{equation}
is the bispectrum reconstructed from the observed $a_{\ell m}$, and
\begin{equation}
F \equiv \sum_{\ell_1 \ell_2 \ell_3} (-1)^{\ell_1 + \ell_2 + \ell_3}
\frac{\left( B_{\ell_1 \ell_2 \ell_3}^{\rm th} \right)^2 }{6 C_{\ell_1} C_{\ell_2} C_{\ell_3}} \label{eq:fish}
\end{equation}
is the Fisher matrix. The bestfit value and the error bar on $f_{\rm NL}$ is obtained computing $\hat{f}_{\rm NL}$ from the observed data and the simulation maps, respectively. The straightforward computation of Eq.~\eqref{eq:estimator_general} is not a wise manner since the $\mathcal{O}(\ell_{\rm max}^5)$ numerical operation required here is quite time-consuming and practically unfeasible for $\ell_{\rm max} \gtrsim 100$.

To reduce the computational cost, factorizing the summation as $\sum_{\ell_1 \ell_2 \ell_3} f(\ell_1, \ell_2, \ell_3) = [\sum_{\ell_1} a(\ell_1) ] [\sum_{\ell_2} b(\ell_2) ] [\sum_{\ell_3} c(\ell_3) ]$ is a tested approach.%
\footnote{
  See Ref.~\citep{Bucher:2015ura} for another approach by binning the $\ell$-space domain.
}
This idea was implemented in Ref.~\citep{Komatsu:2003iq} for the first time, enabling the analysis for originally separable templates such as local, equilateral and orthogonal NGs. Later, it was generalized by developing the modal decomposition technique of any originally non-separable $B_{\ell_1 \ell_2 \ell_3}^{\rm th}$ \citep{Fergusson:2009nv,Fergusson:2010dm}. The original version of the modal methodology is for the analysis of even~$\ell_1 + \ell_2 + \ell_3$ domain, while the spin-weighted version developed recently \citep{Shiraishi:2014roa,Shiraishi:2014ila} can also cover odd~$\ell_1 + \ell_2 + \ell_3$ domain. As seen in Sec.~\ref{sec:CMB_bis}, the tensor-mode CMB bispectrum is generally non-separable and does not always vanish in the odd~$\ell_1 + \ell_2 + \ell_3$ domain. In this sense, the modal methodology is an indispensable tool for the tensor NG search. In fact, except for the results obtained bruteforcing the $\ell$-space summation with $\ell_{\rm max} = 100$ \citep{Shiraishi:2013wua}, all observational constraints reported so far have been obtained by it \citep{Shiraishi:2014ila,Ade:2015ava,Ade:2015cva,Shiraishi:2017yrq,Akrami:2019izv}. In what follows, for simplicity, we explain the modal methodology for the auto bispectra ($TTT$, $EEE$, and $BBB$) based on Refs.~\citep{Fergusson:2009nv,Fergusson:2010dm,Shiraishi:2014roa,Shiraishi:2014ila}, while one can deal with the cross bispectra ($TTE$, $TTB$, $TEE$, $TEB$, $TBB$, $EEB$, and $EBB$) in the very similar way \citep{Fergusson:2014gea,Shiraishi:2019exr}.

Let us introduce the reduced bispectrum according to:
\begin{equation}
B_{\ell_1 \ell_2 \ell_3} 
\equiv h_{\ell_1 \ell_2 \ell_3}^{\{xyz\}} b_{\ell_1 \ell_2 \ell_3} , 
\end{equation}
where, 
\begin{equation}
  h_{\ell_1 \ell_2 \ell_3}^{\{xyz\}} \equiv \frac{1}{6} h_{\ell_1 \ell_2 \ell_3}^{x~y~z} 
+ {5~\rm perm~in~} x, y, z .
\end{equation}
Here and hereinafter, we follow the notation for the symmetric operation $A^{\{x}A^{y}A^{z\}} \equiv \frac{1}{6}A^{x}A^y A^z + {\rm 5~perm~in~}x,y,z$. The spinned weight $(x,y,z)$ is fixed to be, e.g., $(0,0,0)$ and $(1, 1, -2)$ for even and odd $\ell_1 + \ell_2 + \ell_3$ analysis, respectively. The reduced bispectrum of the theoretical template is decomposed in the odd/even $\ell_1 + \ell_2 + \ell_3$ domain separately, according to:
\begin{equation}
\frac{v_{\ell_1} v_{\ell_2} v_{\ell_3}}{\sigma \sqrt{C_{\ell_1} C_{\ell_2} C_{\ell_3}} } b_{\ell_1 \ell_2 \ell_3}^{{\rm th} (o/e)} 
= \sum_{ijk} \alpha_{ijk}^{(o/e)} Q_{ijk}(\ell_1, \ell_2, \ell_3), \label{eq:Q_decomp}
\end{equation}
where the separable modal basis $Q_{ijk}$ is composed of the products of the eigenfunctions $q_i(\ell) \in \mathbb{R} $ as:
\begin{eqnarray}
  Q_{ijk}(\ell_1, \ell_2, \ell_3) 
&\equiv& q_{\{i}(\ell_1) q_{j}(\ell_2) q_{k\}}(\ell_3) \nonumber \\ 
&=& \frac{1}{6} q_i(\ell_1) q_j(\ell_2) q_k(\ell_3) + 5~{\rm perm~in~}i,j,k .
\end{eqnarray}
The $\sigma$ factor, defined by: 
\begin{equation}
\sigma \equiv
\begin{cases}
1 & :\ell_1 + \ell_2 + \ell_3 = {\rm even} \\
i & :\ell_1 + \ell_2 + \ell_3 = {\rm odd}
\end{cases} ,
\end{equation}
makes the L.H.S of Eq.~\eqref{eq:Q_decomp} real, so the modal coefficients $\alpha_{ijk}^{(o/e)}$ are always real. The convergence speed of the modal decomposition relies on the choice of the $v_\ell$ weighting and $q_i(\ell)$. Employing the $Q_{ijk}$ templates, we define a matrix according to:
\begin{equation}
  \gamma_{np}^{(o/e)} \equiv \Braket{Q_n(\ell_1, \ell_2, \ell_3) \, Q_p (\ell_1, \ell_2, \ell_3)}_{o/e} ,
\end{equation}
where the triples $ijk$ in $Q_{ijk}$ are labeled by means of a single index $n$,
and the bracket $\braket{\cdots}_{o/e}$ denotes the summation in terms of the odd/even $\ell_1 + \ell_2 + \ell_3$ domain as:
\begin{equation}
  \Braket{F_{\ell_1 \ell_2 \ell_3}}_{o/e} \equiv 
  \sum_{\ell_1 + \ell_2 + \ell_3 = {\rm odd / even}} 
\left( \frac{h_{\ell_1 \ell_2 \ell_3}^{\{xyz\}}}{v_{\ell_1} v_{\ell_2} v_{\ell_3}} \right)^2 F_{\ell_1 \ell_2 \ell_3} .
\end{equation}
Then, the modal coefficients are computed according to the inverse operation:
\begin{equation}
\alpha_n^{(o/e)} = \sum_p \left( \gamma^{(o/e)} \right)_{np}^{-1} 
\Braket{ \frac{v_{\ell_1} v_{\ell_2} v_{\ell_3} b_{\ell_1 \ell_2 \ell_3}^{{\rm th} (o/e)}}{\sigma \sqrt{C_{\ell_1} C_{\ell_2} C_{\ell_3}}} Q_p(\ell_1,\ell_2,\ell_3) }_{o/e} .
\end{equation}

Using Eq.~\eqref{eq:Q_decomp} and the identity:
\begin{equation}
  h_{l_1 l_2 l_3}^{s_1 s_2 s_3}
  \left(
  \begin{array}{ccc}
    l_1 & l_2 & l_3 \\
    m_1 & m_2 & m_3
  \end{array}
  \right)
  = \int d^2 \hat{n} \,
  {}_{-s_1}Y_{l_1 m_1}(\hat{n}) \,
  {}_{-s_2}Y_{l_2 m_2}(\hat{n}) \,
  {}_{-s_3}Y_{l_3 m_3}(\hat{n}) ,
\end{equation}
the sum over $\ell_1$, $\ell_2$, and $\ell_3$ in the estimator \eqref{eq:estimator_general} is rewritten into the sum over finite eigenmodes as:
\begin{equation}
  \hat{f}_{\rm NL} = \frac{\sum_n \alpha_n^{(o)} \beta_n^{(o)} + \sum_n \alpha_n^{(e)} \beta_n^{(e)}}{\sum_{np} \alpha_n^{(o)} \gamma_{np}^{(o)} \alpha_p^{(o)} + \sum_{np} \alpha_n^{(e)} \gamma_{np}^{(e)} \alpha_p^{(e)}} , \label{eq:estimator_modal}
\end{equation}
where the $\beta_n^{(o/e)}$ coefficients, reading: 
\begin{eqnarray}
\beta_{ijk}^{(o/e)}  &=&
\frac{1}{\sigma} \int d^2 \hat{n} \sum_{a+b+c = o/e} 
\left[ {}_{\{-x}M_{\{i}^{(a)}(\hat{n}) \, {}_{-y}M_{j}^{ (b)}(\hat{n}) \, {}_{-z\}}M_{k\}}^{(c)}(\hat{n}) \right. \nonumber \\
  &&\left. \qquad\qquad\qquad 
  - 3 \Braket{{}_{\{-x}M_{\{i}^{(a)}(\hat{n}) \, {}_{-y}M_{j}^{(b)}(\hat{n})} {}_{-z\}}M_{k\}}^{(c)}(\hat{n}) \right], 
\end{eqnarray}
are computed from the filtered maps: 
\begin{equation}
{}_x M_{i}^{(o/e)}(\hat{n}) \equiv \sum_{\ell = {\rm odd / even}} \sum_m q_i(\ell)
\frac{a_{\ell m}}{v_\ell \sqrt{C_\ell}} \, {}_{x}Y_{\ell m}(\hat{n}) .
\end{equation}
The most time-consuming task is the computation of $\braket{\cdots}_{o/e}$ in $\alpha_n^{(o/e)}$, however, it requires, at most, $\mathcal{O}(\ell_{\rm max}^3)$ operations (assuming that the modal decomposition converges within a reasonable time), and hence the separable estimator \eqref{eq:estimator_modal} makes the tensor bispectrum analysis feasible.


\subsection{Current observational limits} \label{subsec:limits}

\begin{table}[t]
\begin{center}
\begin{tabular}{|c||c|c|c|} \hline
  & $f_{\rm NL}^{ttt, \rm eq} $ & $f_{\rm NL}^{ttt, \rm sq} $ & $f_{\rm NL}^{tss, \rm sq}$ \\ \hline\hline
  WMAP $T$ only & \,\,\,$600 \pm 1500$ \citep{Shiraishi:2014ila} & $220 \pm 170$ \citep{Shiraishi:2013wua} & $84 \pm 49$ \citep{Shiraishi:2017yrq} \\ 
  {\it Planck} $T$ only & \,\,\,$600 \pm 1600$ \citep{Akrami:2019izv} & $290 \pm 180$ \citep{Ade:2015cva} & -- \\
  {\it Planck} $E$ only & $2900 \pm 6700$ \citep{Akrami:2019izv} & -- & -- \\   
  {\it Planck} $T+E$ & \,\,\,$800 \pm 1100$ \citep{Akrami:2019izv} & -- & -- \\  \hline
  \end{tabular}
\end{center}
\caption{CMB constraints on three tensor non-linearity parameters: $f_{\rm NL}^{ttt, \rm eq} $ \eqref{eq:fnl_def_ttt_eq}, $f_{\rm NL}^{ttt, \rm sq}$ \eqref{eq:fnl_def_ttt_sq}, and $f_{\rm NL}^{tss, \rm sq}$ \eqref{eq:fnl_def_tss_sq}. The central values and $1\sigma$ errors estimated from the WMAP temperature data only (first row), the {\it Planck} temperature data only (second one), the {\it Planck} $E$-mode polarization data only (third one), and the {\it Planck} temperature and $E$-mode polarization data jointly (fourth one) are shown. The results of $f_{\rm NL}^{ttt, \rm eq}$ are obtained by analyzing both even and odd $\ell_1 + \ell_2 + \ell_3$ multipoles, while those of $f_{\rm NL}^{ttt, \rm sq}$ and $f_{\rm NL}^{tss, \rm sq}$ come from only even $\ell_1 + \ell_2 + \ell_3$ modes.} 
\label{tab:fNL}
\end{table}

Table~\ref{tab:fNL} summarizes the current CMB limits on $f_{\rm NL}^{ttt, \rm eq} $, $f_{\rm NL}^{ttt, \rm sq}$, and $f_{\rm NL}^{tss, \rm sq}$ obtained from the WMAP and ${\it Planck}$ maps.%
\footnote{See Ref.~\citep{Shiraishi:2014ila} for the CMB limits on a few other shapes.}
Regarding $f_{\rm NL}^{ttt, \rm eq}$ and $f_{\rm NL}^{ttt, \rm sq}$, the consistency between the WMAP $T$ only results and {\it Planck} $T$ only ones is confirmed there. Note that the unimproved constraints are expected because, as inferred from Fig.~\ref{fig:3D}, $B_{\ell_1 \ell_2 \ell_3}^{\rm th}$ decays rapidly for $\ell \gtrsim 100$ and hence the estimator sum \eqref{eq:estimator_general} saturates well below the WMAP resolution ($\ell_{\rm max} = 500$). In contrast, for $f_{\rm NL}^{tss, \rm sq}$ the current constraint from WMAP is expected to be much improved by {\it Planck} since the squeezed-limit signal survives even for high $\ell$ \citep{Shiraishi:2010kd,Domenech:2017kno}. We also confirm the improvement of the constraint on $f_{\rm NL}^{ttt, \rm eq}$ by analyzing the temperature and $E$-mode polarization data jointly.

As seen in Table~\ref{tab:fNL}, no significant deviation from Gaussianity has been found so far.


\subsection{Future prospects} \label{subsec:future}

\begin{figure*}[t]
  \begin{tabular}{ccc} 
   \begin{minipage}{0.49\hsize}
     \begin{center}
       \includegraphics[width=1\textwidth]{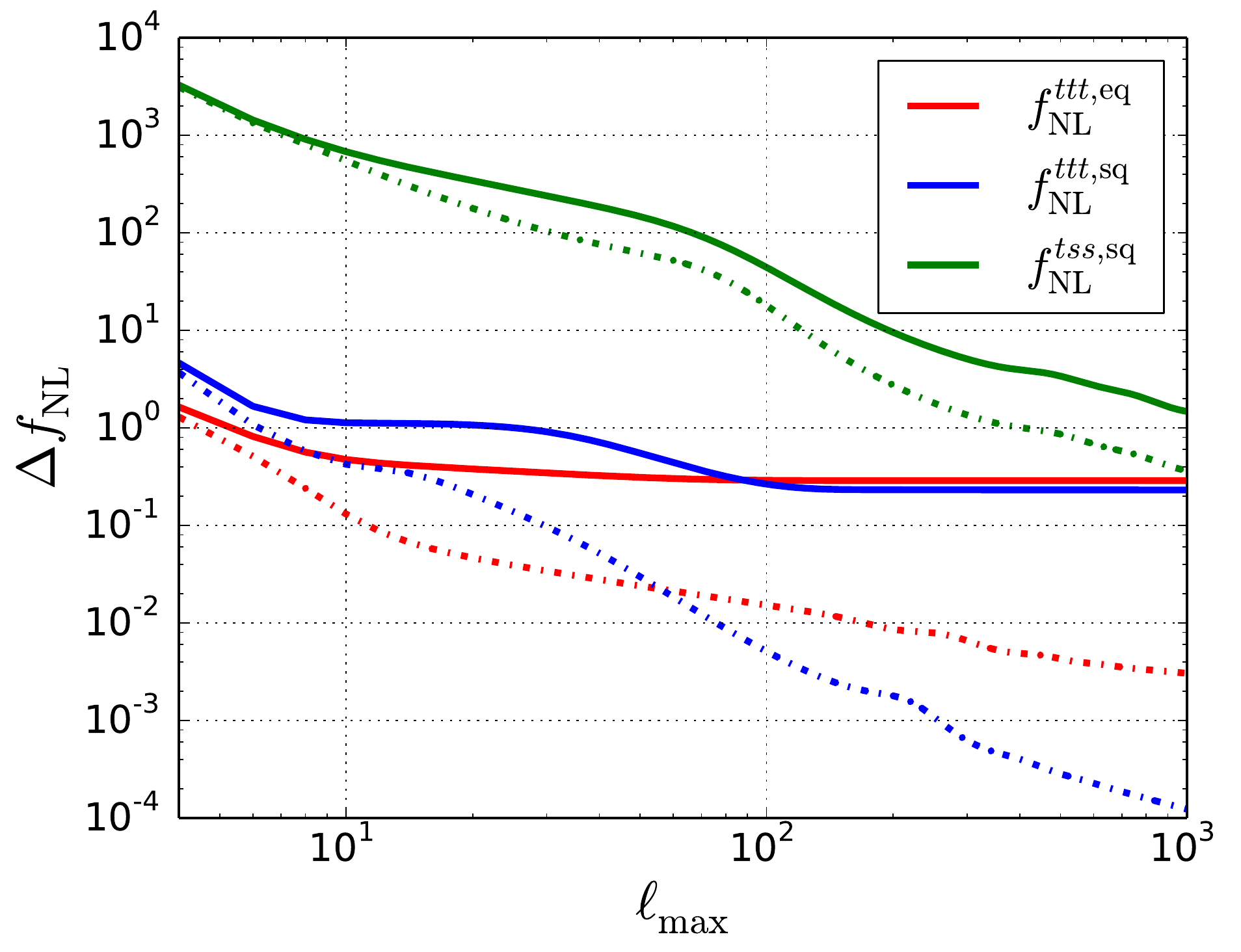}
  \end{center}
   \end{minipage}
   \begin{minipage}{0.49\hsize}
     \begin{center}
       \includegraphics[width=1\textwidth]{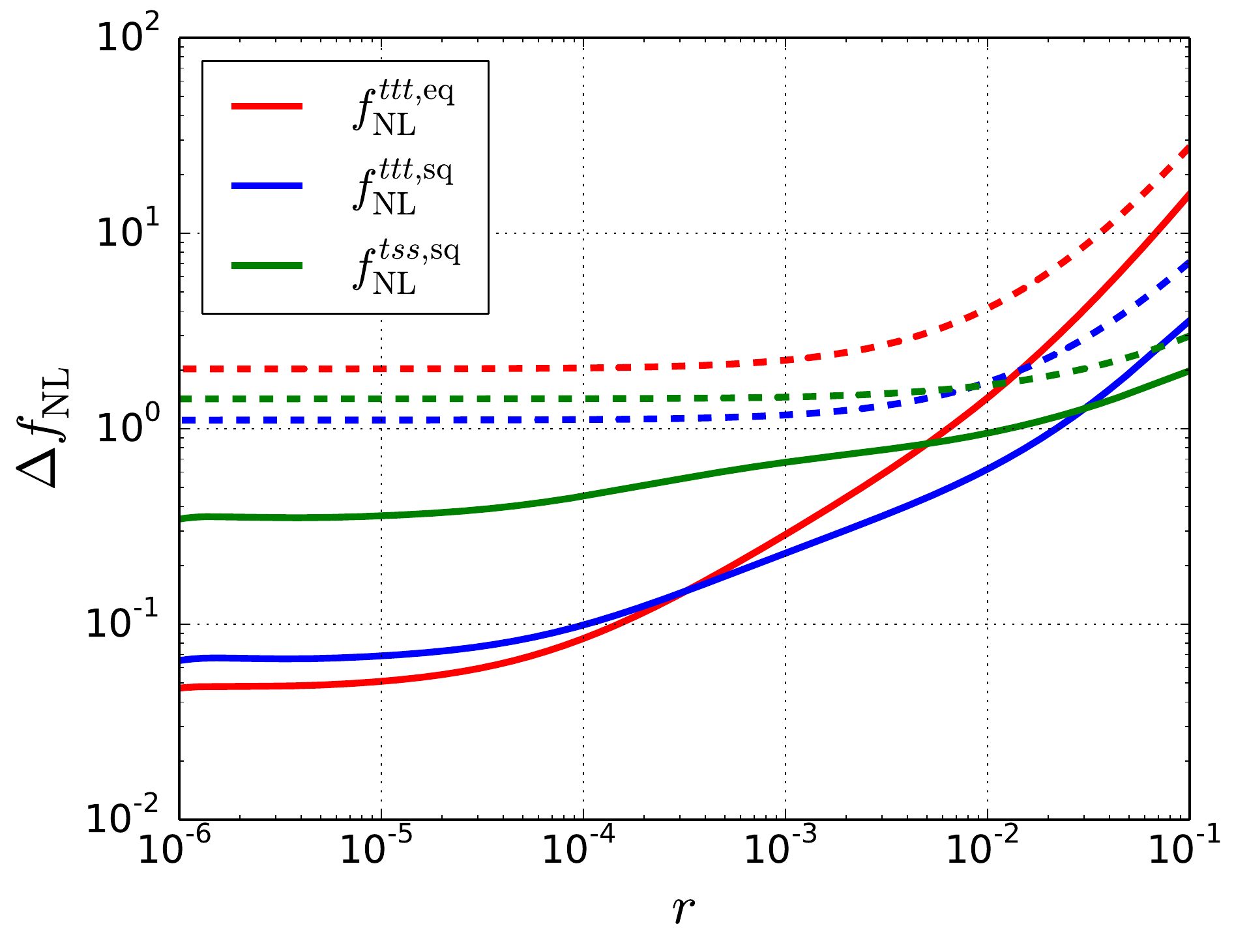}
  \end{center}
   \end{minipage}
  \end{tabular}
  \caption{Expected $1\sigma$ errors: $\Delta f_{\rm NL}^{ttt, \rm eq}$ (red lines) and $\Delta f_{\rm NL}^{ttt, \rm sq}$ (blue lines) from $BBB$, and $\Delta f_{\rm NL}^{tss, \rm sq}$ (green lines) from $BTT$, as a function of the maximum multipole number $\ell_{\rm max}$ (left panel) and the tensor-to-scalar ratio $r$ (right panel). The left panel is depicted adopting $r = 10^{-3}$, and $\Delta f_{\rm NL}^{tss, \rm sq}$ in the right panel is computed with $\ell_{\rm max} = 2000$. The linestyle discriminates the cleanliness level of the $B$-mode data: a perfectly-delensed and noiseless full-sky case (dot-dashed lines in left panel), a non-delensed and noiseless full-sky one (solid lines in both panels), and a LiteBIRD-like realistic one (dashed lines in right panel).}
\label{fig:error}
\end{figure*}

Finally, we discuss future prospects of detecting these tensor NGs by adding $B$-mode polarization to the data analysis. Here, we evaluate expected $1\sigma$ errors $\Delta f_{\rm NL}^{ttt, \rm eq}$ and $\Delta f_{\rm NL}^{ttt, \rm sq}$ from $BBB$, and $\Delta f_{\rm NL}^{tss, \rm sq}$ from $BTT$, through the computation of the Fisher matrix. The covariance matrices [corresponding to the denominator of Eq.~\eqref{eq:fish}] are given by $C_{\ell_1}^{BB} C_{\ell_2}^{BB} C_{\ell_3}^{BB}$ and $C_{\ell_1}^{BB} C_{\ell_2}^{TT} C_{\ell_3}^{TT}$, respectively. Note that these are estimated under the diagonal covariance matrix approximation. The $B$-mode power spectrum is computed by summing up the primordial contribution parametrized by the tensor-to-scalar ratio $r$, the lensing $B$-mode one and the experimental noise spectrum as $C_\ell^{BB} = C_\ell^{\rm prim}(r) + C_\ell^{\rm lens}  + N_\ell$. Then, let us examine three cleanliness levels of the $B$-mode data: a perfectly-delensed and noiseless full-sky case (i.e., $C_\ell^{\rm lens}  = N_\ell = 0$), a non-delensed and noiseless full-sky one (i.e., $N_\ell = 0$), and a non-delensed realistic case where experimental uncertainties due to beam, noise, mask and residual foreground in LiteBIRD \citep{Hazumi:2012aa,Matsumura:2013aja,2016JLTP..tmp..169M} are considered.%
\footnote{
  In the third case, we compute $N_\ell$ by taking into account the contamination due to residual foregrounds in the same manner as Ref.~\citep{Shiraishi:2016yun}; namely, we assume that foregrounds due to galactic dust emission and synchrotron radiation are subtracted using 9 channels (corresponding to 40-89~GHz and 280-402~GHz) and therefore reduce to the 2\% level in CMB maps. Any higher-order contribution is not considered for simplicity.
}
The first example provides the theoretical limits. In this case, the size of the covariance matrix is determined by $r$ alone and, accordingly, the errors simply scale like:
\begin{equation}
  \Delta f_{\rm NL}^{ttt, \rm eq} \propto r^{3/2} , \ \ \
  \Delta f_{\rm NL}^{ttt, \rm sq} \propto r^{3/2} , \ \ \
  \Delta f_{\rm NL}^{tss, \rm sq} \propto r^{1/2} . \label{eq:error_CVL} 
\end{equation}
Comparing this with Eq.~\eqref{eq:fnl_ttt_eq_pseudo} or Eq.~\eqref{eq:fnl_tss_sq_massive}, one can see that the error on a model parameter of the axion model, $\Omega_A^{-1}$, or that of the massive gravity model, $\lambda_{sst}$, is proportional to $r^{-1/2}$. 

The left panel of Fig.~\ref{fig:error} describes the $\ell_{\rm max}$ dependence of $\Delta f_{\rm NL}$ in the perfectly-delensed and non-delensed cases with $r = 10^{-3}$ and $N_\ell = 0$. From the former results (corresponding to the dot-dashed lines), the sensitivity improvement by increasing $\ell_{\rm max}$ is confirmed. From the latter results (corresponding to the solid lines), it is visually apparent that $\Delta f_{\rm NL}^{ttt, \rm eq}$ and $\Delta f_{\rm NL}^{ttt, \rm sq}$ immediately saturate because $C_\ell^{\rm lens}$ dominates over $C_{\ell}^{BB}$ for $\ell > \mathcal{O}(10)$. In contrast, $\Delta f_{\rm NL}^{ttt, \rm sq}$ is free from such a degradation since the dominant signal in the Fisher matrix comes from large-scale $B$ modes, more precisely, long-wavelength $B$ and short-wavelength $T$ squeezed configurations ($\ell_1^B \ll \ell_2^T \sim \ell_3^T$) \citep{Domenech:2017kno,Meerburg:2016ecv}. However, of course, $\Delta f_{\rm NL}^{ttt, \rm sq}$ would also saturate at very small scales, i.e., $\ell^T \gtrsim 3000$, where the scalar-mode lensing contamination dominates $C_{\ell}^{TT}$. Higher-order lensing contributions introduce non-vanishing off-diagonal components in the covariance matrix; thus, the above simple Fisher matrix computation would no longer be credible. In the left panel of Fig.~\ref{fig:error}, the results of the LiteBIRD-like experiment are not shown, however, they have similar $\ell_{\rm max}$ scalings and the slightly larger overall sizes in comparison with the no-delensed and noiseless full-sky results.

In the right panel of Fig.~\ref{fig:error}, the $r$ dependence of $\Delta f_{\rm NL}$ for the no-delensed and noiseless case (corresponding to the solid lines) and a LiteBIRD-like experiment (corresponding to the dashed lines) is presented. For large $r$, $C_\ell^{\rm prim}$ dominates over $C_\ell^{BB}$ within a wide range of $\ell$ and hence the $r$ dependence reaches the ideal case \eqref{eq:error_CVL}. On the other hand, for decreasing $r$, $C_\ell^{\rm lens}$ or $N_\ell$ starts to dominate and errors converge. We find that LiteBIRD could measure an $\mathcal{O}(1)$ signal of $f_{\rm NL}^{ttt, \rm eq}$, $f_{\rm NL}^{ttt, \rm sq}$, or $f_{\rm NL}^{tss, \rm sq}$.%
\footnote{
Comparable detectability is expected in the other next-generation CMB experiments such as CMB-S4 \citep{Abazajian:2016yjj} and CORE \citep{Delabrouille:2017rct}.
}
If detected, as discussed in Sec.~\ref{sec:models}, we will have compelling evidence for a deviation from Einstein gravity or the existence of extra source fields, which will help establish an improved picture of the early Universe.

Finally, it is worth mentioning impacts of late-time secondary bispectra. Via gravitational lensing, primordial temperature and $E$-mode polarization fields induce secondary $BTT$ signals \citep{Hu:2000ee,Lewis:2011fk}. Fortunately, this has a very weak correlation with the $f_{\rm NL}^{tss, \rm sq}$ bispectrum template \cite{Domenech:2017kno}. Similarly, secondary $BBB$ is also induced, while this is subdominant compared to primordial $BBB$ at interesting $f_{\rm NL}^{ttt, \rm eq}$ or $f_{\rm NL}^{ttt, \rm sq}$ \citep{Shiraishi:2016yun}. Besides of these leading-order contributions, higher-order ones exist due to post-Born lensing \citep{Pratten:2016dsm,Marozzi:2016uob}. Secondary polarized bispectra can also be generated via the Sunyaev-Zel'dovich effect \citep{Sunyaev:1972eq} and extragalactic foreground \citep{Jung:2018rgf,Coulton:2019bnz}. These impacts are still non-trivial and should therefore be solved for more precise discussions.

\section{Conclusions} \label{sec:conclusions}

The search for tensor NG is theoretically well-motivated, and the methodology for testing with the CMB temperature and E/B-mode polarization data has already been established. Observational constraints on some templates from the WMAP temperature and the {\it Planck} temperature and $E$-mode polarization data do exist. Any significant signal has not yet been discovered, implying the smallness of tensor NGs. There is still a chance of approaching such a small signal by including $B$-mode polarization in the data analysis. As we have found, an $\mathcal{O}(1)$ signal of $f_{\rm NL}^{ttt, \rm eq}$, $f_{\rm NL}^{ttt, \rm sq}$, or $f_{\rm NL}^{tss, \rm sq}$ would be detectable in LiteBIRD; current constraints are expected to be updated in the next decade.

Throughout this paper, we have focused mainly on scale-invariant cases, while specific scale dependences can be generated for tensor NGs depending on the shapes of the inflationary potentials and non-linear interactions. The detectability is then enhanced at specific multipoles \citep{Shiraishi:2016yun} or outside the CMB scales. In the latter case, the information from other probes such as an intereferometric GW survey (e.g., \citep{Bartolo:2018qqn,Tsuneto:2018tif}), the galaxy one (e.g., statistical anisotropy in the galaxy power spectrum \citep{Jeong:2012df}) and the 21-cm line one also becomes indispensable. A more comprehensive analysis based on multi-wavelength observations remains an interesting and important future issue.



\section*{Acknowledgments}

I thank the editors of ``Status and Prospects of Cosmic Microwave Background Analysis'' and especially, Maurizio Tomasi and Michele Liguori for their invitation to contribute to this special issue and for all their support. I also appreciate the careful reviews of the manuscripts by the referees. This work was supported by the JSPS Grant-in-Aid for Research Activity Start-up Grant Number 17H07319 and the JSPS Grant-in-Aid for Early-Career Scientists Grant Number 19K14718. Numerical computations were in part carried out on Cray XC50 at the Center for Computational Astrophysics, National Astronomical Observatory of Japan.



\bibliographystyle{apsrev4-1}
\bibliography{paper}

\begin{thebibliography}{73}%
\makeatletter
\providecommand \@ifxundefined [1]{%
 \@ifx{#1\undefined}
}%
\providecommand \@ifnum [1]{%
 \ifnum #1\expandafter \@firstoftwo
 \else \expandafter \@secondoftwo
 \fi
}%
\providecommand \@ifx [1]{%
 \ifx #1\expandafter \@firstoftwo
 \else \expandafter \@secondoftwo
 \fi
}%
\providecommand \natexlab [1]{#1}%
\providecommand \enquote  [1]{``#1''}%
\providecommand \bibnamefont  [1]{#1}%
\providecommand \bibfnamefont [1]{#1}%
\providecommand \citenamefont [1]{#1}%
\providecommand \href@noop [0]{\@secondoftwo}%
\providecommand \href [0]{\begingroup \@sanitize@url \@href}%
\providecommand \@href[1]{\@@startlink{#1}\@@href}%
\providecommand \@@href[1]{\endgroup#1\@@endlink}%
\providecommand \@sanitize@url [0]{\catcode `\\12\catcode `\$12\catcode
  `\&12\catcode `\#12\catcode `\^12\catcode `\_12\catcode `\%12\relax}%
\providecommand \@@startlink[1]{}%
\providecommand \@@endlink[0]{}%
\providecommand \url  [0]{\begingroup\@sanitize@url \@url }%
\providecommand \@url [1]{\endgroup\@href {#1}{\urlprefix }}%
\providecommand \urlprefix  [0]{URL }%
\providecommand \Eprint [0]{\href }%
\providecommand \doibase [0]{http://dx.doi.org/}%
\providecommand \selectlanguage [0]{\@gobble}%
\providecommand \bibinfo  [0]{\@secondoftwo}%
\providecommand \bibfield  [0]{\@secondoftwo}%
\providecommand \translation [1]{[#1]}%
\providecommand \BibitemOpen [0]{}%
\providecommand \bibitemStop [0]{}%
\providecommand \bibitemNoStop [0]{.\EOS\space}%
\providecommand \EOS [0]{\spacefactor3000\relax}%
\providecommand \BibitemShut  [1]{\csname bibitem#1\endcsname}%
\let\auto@bib@innerbib\@empty
\bibitem [{\citenamefont {Aghanim}\ \emph {et~al.}(2018)\citenamefont {Aghanim}
  \emph {et~al.}}]{Aghanim:2018eyx}%
  \BibitemOpen
  \bibfield  {author} {\bibinfo {author} {\bibfnamefont {N.}~\bibnamefont
  {Aghanim}} \emph {et~al.} (\bibinfo {collaboration} {Planck}),\ }\href@noop
  {} {\  (\bibinfo {year} {2018})},\ \Eprint {http://arxiv.org/abs/1807.06209}
  {arXiv:1807.06209 [astro-ph.CO]} \BibitemShut {NoStop}%
\bibitem [{\citenamefont {Akrami}\ \emph {et~al.}(2018)\citenamefont {Akrami}
  \emph {et~al.}}]{Akrami:2018odb}%
  \BibitemOpen
  \bibfield  {author} {\bibinfo {author} {\bibfnamefont {Y.}~\bibnamefont
  {Akrami}} \emph {et~al.} (\bibinfo {collaboration} {Planck}),\ }\href@noop {}
  {\  (\bibinfo {year} {2018})},\ \Eprint {http://arxiv.org/abs/1807.06211}
  {arXiv:1807.06211 [astro-ph.CO]} \BibitemShut {NoStop}%
\bibitem [{\citenamefont {Kamionkowski}\ and\ \citenamefont
  {Kovetz}(2016)}]{Kamionkowski:2015yta}%
  \BibitemOpen
  \bibfield  {author} {\bibinfo {author} {\bibfnamefont {M.}~\bibnamefont
  {Kamionkowski}}\ and\ \bibinfo {author} {\bibfnamefont {E.~D.}\ \bibnamefont
  {Kovetz}},\ }\href {\doibase 10.1146/annurev-astro-081915-023433} {\bibfield
  {journal} {\bibinfo  {journal} {Ann. Rev. Astron. Astrophys.}\ }\textbf
  {\bibinfo {volume} {54}},\ \bibinfo {pages} {227} (\bibinfo {year} {2016})},\
  \Eprint {http://arxiv.org/abs/1510.06042} {arXiv:1510.06042 [astro-ph.CO]}
  \BibitemShut {NoStop}%
\bibitem [{\citenamefont {Guzzetti}\ \emph {et~al.}(2016)\citenamefont
  {Guzzetti}, \citenamefont {Bartolo}, \citenamefont {Liguori},\ and\
  \citenamefont {Matarrese}}]{Guzzetti:2016mkm}%
  \BibitemOpen
  \bibfield  {author} {\bibinfo {author} {\bibfnamefont {M.~C.}\ \bibnamefont
  {Guzzetti}}, \bibinfo {author} {\bibfnamefont {N.}~\bibnamefont {Bartolo}},
  \bibinfo {author} {\bibfnamefont {M.}~\bibnamefont {Liguori}}, \ and\
  \bibinfo {author} {\bibfnamefont {S.}~\bibnamefont {Matarrese}},\ }\href
  {\doibase 10.1393/ncr/i2016-10127-1} {\bibfield  {journal} {\bibinfo
  {journal} {Riv. Nuovo Cim.}\ }\textbf {\bibinfo {volume} {39}},\ \bibinfo
  {pages} {399} (\bibinfo {year} {2016})},\ \Eprint
  {http://arxiv.org/abs/1605.01615} {arXiv:1605.01615 [astro-ph.CO]}
  \BibitemShut {NoStop}%
\bibitem [{\citenamefont {Acquaviva}\ \emph {et~al.}(2003)\citenamefont
  {Acquaviva}, \citenamefont {Bartolo}, \citenamefont {Matarrese},\ and\
  \citenamefont {Riotto}}]{Acquaviva:2002ud}%
  \BibitemOpen
  \bibfield  {author} {\bibinfo {author} {\bibfnamefont {V.}~\bibnamefont
  {Acquaviva}}, \bibinfo {author} {\bibfnamefont {N.}~\bibnamefont {Bartolo}},
  \bibinfo {author} {\bibfnamefont {S.}~\bibnamefont {Matarrese}}, \ and\
  \bibinfo {author} {\bibfnamefont {A.}~\bibnamefont {Riotto}},\ }\href
  {\doibase 10.1016/S0550-3213(03)00550-9} {\bibfield  {journal} {\bibinfo
  {journal} {Nucl. Phys.}\ }\textbf {\bibinfo {volume} {B667}},\ \bibinfo
  {pages} {119} (\bibinfo {year} {2003})},\ \Eprint
  {http://arxiv.org/abs/astro-ph/0209156} {arXiv:astro-ph/0209156 [astro-ph]}
  \BibitemShut {NoStop}%
\bibitem [{\citenamefont {Maldacena}(2003)}]{Maldacena:2002vr}%
  \BibitemOpen
  \bibfield  {author} {\bibinfo {author} {\bibfnamefont {J.~M.}\ \bibnamefont
  {Maldacena}},\ }\href {\doibase 10.1088/1126-6708/2003/05/013} {\bibfield
  {journal} {\bibinfo  {journal} {JHEP}\ }\textbf {\bibinfo {volume} {05}},\
  \bibinfo {pages} {013} (\bibinfo {year} {2003})},\ \Eprint
  {http://arxiv.org/abs/astro-ph/0210603} {arXiv:astro-ph/0210603 [astro-ph]}
  \BibitemShut {NoStop}%
\bibitem [{\citenamefont {Namba}\ \emph {et~al.}(2016)\citenamefont {Namba},
  \citenamefont {Peloso}, \citenamefont {Shiraishi}, \citenamefont {Sorbo},\
  and\ \citenamefont {Unal}}]{Namba:2015gja}%
  \BibitemOpen
  \bibfield  {author} {\bibinfo {author} {\bibfnamefont {R.}~\bibnamefont
  {Namba}}, \bibinfo {author} {\bibfnamefont {M.}~\bibnamefont {Peloso}},
  \bibinfo {author} {\bibfnamefont {M.}~\bibnamefont {Shiraishi}}, \bibinfo
  {author} {\bibfnamefont {L.}~\bibnamefont {Sorbo}}, \ and\ \bibinfo {author}
  {\bibfnamefont {C.}~\bibnamefont {Unal}},\ }\href {\doibase
  10.1088/1475-7516/2016/01/041} {\bibfield  {journal} {\bibinfo  {journal}
  {JCAP}\ }\textbf {\bibinfo {volume} {1601}},\ \bibinfo {pages} {041}
  (\bibinfo {year} {2016})},\ \Eprint {http://arxiv.org/abs/1509.07521}
  {arXiv:1509.07521 [astro-ph.CO]} \BibitemShut {NoStop}%
\bibitem [{\citenamefont {Agrawal}\ \emph
  {et~al.}(2018{\natexlab{a}})\citenamefont {Agrawal}, \citenamefont {Fujita},\
  and\ \citenamefont {Komatsu}}]{Agrawal:2017awz}%
  \BibitemOpen
  \bibfield  {author} {\bibinfo {author} {\bibfnamefont {A.}~\bibnamefont
  {Agrawal}}, \bibinfo {author} {\bibfnamefont {T.}~\bibnamefont {Fujita}}, \
  and\ \bibinfo {author} {\bibfnamefont {E.}~\bibnamefont {Komatsu}},\ }\href
  {\doibase 10.1103/PhysRevD.97.103526} {\bibfield  {journal} {\bibinfo
  {journal} {Phys. Rev.}\ }\textbf {\bibinfo {volume} {D97}},\ \bibinfo {pages}
  {103526} (\bibinfo {year} {2018}{\natexlab{a}})},\ \Eprint
  {http://arxiv.org/abs/1707.03023} {arXiv:1707.03023 [astro-ph.CO]}
  \BibitemShut {NoStop}%
\bibitem [{\citenamefont {Dimastrogiovanni}\ \emph {et~al.}(2019)\citenamefont
  {Dimastrogiovanni}, \citenamefont {Fasiello}, \citenamefont {Tasinato},\ and\
  \citenamefont {Wands}}]{Dimastrogiovanni:2018gkl}%
  \BibitemOpen
  \bibfield  {author} {\bibinfo {author} {\bibfnamefont {E.}~\bibnamefont
  {Dimastrogiovanni}}, \bibinfo {author} {\bibfnamefont {M.}~\bibnamefont
  {Fasiello}}, \bibinfo {author} {\bibfnamefont {G.}~\bibnamefont {Tasinato}},
  \ and\ \bibinfo {author} {\bibfnamefont {D.}~\bibnamefont {Wands}},\ }\href
  {\doibase 10.1088/1475-7516/2019/02/008} {\bibfield  {journal} {\bibinfo
  {journal} {JCAP}\ }\textbf {\bibinfo {volume} {1902}},\ \bibinfo {pages}
  {008} (\bibinfo {year} {2019})},\ \Eprint {http://arxiv.org/abs/1810.08866}
  {arXiv:1810.08866 [astro-ph.CO]} \BibitemShut {NoStop}%
\bibitem [{\citenamefont {Goon}\ \emph {et~al.}(2018)\citenamefont {Goon},
  \citenamefont {Hinterbichler}, \citenamefont {Joyce},\ and\ \citenamefont
  {Trodden}}]{Goon:2018fyu}%
  \BibitemOpen
  \bibfield  {author} {\bibinfo {author} {\bibfnamefont {G.}~\bibnamefont
  {Goon}}, \bibinfo {author} {\bibfnamefont {K.}~\bibnamefont {Hinterbichler}},
  \bibinfo {author} {\bibfnamefont {A.}~\bibnamefont {Joyce}}, \ and\ \bibinfo
  {author} {\bibfnamefont {M.}~\bibnamefont {Trodden}},\ }\href@noop {} {\
  (\bibinfo {year} {2018})},\ \Eprint {http://arxiv.org/abs/1812.07571}
  {arXiv:1812.07571 [hep-th]} \BibitemShut {NoStop}%
\bibitem [{\citenamefont {Shiraishi}\ \emph
  {et~al.}(2011{\natexlab{a}})\citenamefont {Shiraishi}, \citenamefont {Nitta},
  \citenamefont {Yokoyama}, \citenamefont {Ichiki},\ and\ \citenamefont
  {Takahashi}}]{Shiraishi:2011dh}%
  \BibitemOpen
  \bibfield  {author} {\bibinfo {author} {\bibfnamefont {M.}~\bibnamefont
  {Shiraishi}}, \bibinfo {author} {\bibfnamefont {D.}~\bibnamefont {Nitta}},
  \bibinfo {author} {\bibfnamefont {S.}~\bibnamefont {Yokoyama}}, \bibinfo
  {author} {\bibfnamefont {K.}~\bibnamefont {Ichiki}}, \ and\ \bibinfo {author}
  {\bibfnamefont {K.}~\bibnamefont {Takahashi}},\ }\href {\doibase
  10.1103/PhysRevD.83.123003} {\bibfield  {journal} {\bibinfo  {journal} {Phys.
  Rev.}\ }\textbf {\bibinfo {volume} {D83}},\ \bibinfo {pages} {123003}
  (\bibinfo {year} {2011}{\natexlab{a}})},\ \Eprint
  {http://arxiv.org/abs/1103.4103} {arXiv:1103.4103 [astro-ph.CO]} \BibitemShut
  {NoStop}%
\bibitem [{\citenamefont {Maldacena}\ and\ \citenamefont
  {Pimentel}(2011)}]{Maldacena:2011nz}%
  \BibitemOpen
  \bibfield  {author} {\bibinfo {author} {\bibfnamefont {J.~M.}\ \bibnamefont
  {Maldacena}}\ and\ \bibinfo {author} {\bibfnamefont {G.~L.}\ \bibnamefont
  {Pimentel}},\ }\href {\doibase 10.1007/JHEP09(2011)045} {\bibfield  {journal}
  {\bibinfo  {journal} {JHEP}\ }\textbf {\bibinfo {volume} {09}},\ \bibinfo
  {pages} {045} (\bibinfo {year} {2011})},\ \Eprint
  {http://arxiv.org/abs/1104.2846} {arXiv:1104.2846 [hep-th]} \BibitemShut
  {NoStop}%
\bibitem [{\citenamefont {Gao}\ \emph {et~al.}(2011)\citenamefont {Gao},
  \citenamefont {Kobayashi}, \citenamefont {Yamaguchi},\ and\ \citenamefont
  {Yokoyama}}]{Gao:2011vs}%
  \BibitemOpen
  \bibfield  {author} {\bibinfo {author} {\bibfnamefont {X.}~\bibnamefont
  {Gao}}, \bibinfo {author} {\bibfnamefont {T.}~\bibnamefont {Kobayashi}},
  \bibinfo {author} {\bibfnamefont {M.}~\bibnamefont {Yamaguchi}}, \ and\
  \bibinfo {author} {\bibfnamefont {J.}~\bibnamefont {Yokoyama}},\ }\href
  {\doibase 10.1103/PhysRevLett.107.211301} {\bibfield  {journal} {\bibinfo
  {journal} {Phys. Rev. Lett.}\ }\textbf {\bibinfo {volume} {107}},\ \bibinfo
  {pages} {211301} (\bibinfo {year} {2011})},\ \Eprint
  {http://arxiv.org/abs/1108.3513} {arXiv:1108.3513 [astro-ph.CO]} \BibitemShut
  {NoStop}%
\bibitem [{\citenamefont {Gao}\ \emph {et~al.}(2013)\citenamefont {Gao},
  \citenamefont {Kobayashi}, \citenamefont {Shiraishi}, \citenamefont
  {Yamaguchi}, \citenamefont {Yokoyama},\ and\ \citenamefont
  {Yokoyama}}]{Gao:2012ib}%
  \BibitemOpen
  \bibfield  {author} {\bibinfo {author} {\bibfnamefont {X.}~\bibnamefont
  {Gao}}, \bibinfo {author} {\bibfnamefont {T.}~\bibnamefont {Kobayashi}},
  \bibinfo {author} {\bibfnamefont {M.}~\bibnamefont {Shiraishi}}, \bibinfo
  {author} {\bibfnamefont {M.}~\bibnamefont {Yamaguchi}}, \bibinfo {author}
  {\bibfnamefont {J.}~\bibnamefont {Yokoyama}}, \ and\ \bibinfo {author}
  {\bibfnamefont {S.}~\bibnamefont {Yokoyama}},\ }\href {\doibase
  10.1093/ptep/ptt031} {\bibfield  {journal} {\bibinfo  {journal} {PTEP}\
  }\textbf {\bibinfo {volume} {2013}},\ \bibinfo {pages} {053E03} (\bibinfo
  {year} {2013})},\ \Eprint {http://arxiv.org/abs/1207.0588} {arXiv:1207.0588
  [astro-ph.CO]} \BibitemShut {NoStop}%
\bibitem [{\citenamefont {Akita}\ and\ \citenamefont
  {Kobayashi}(2016)}]{Akita:2015mho}%
  \BibitemOpen
  \bibfield  {author} {\bibinfo {author} {\bibfnamefont {Y.}~\bibnamefont
  {Akita}}\ and\ \bibinfo {author} {\bibfnamefont {T.}~\bibnamefont
  {Kobayashi}},\ }\href {\doibase 10.1103/PhysRevD.93.043519} {\bibfield
  {journal} {\bibinfo  {journal} {Phys. Rev.}\ }\textbf {\bibinfo {volume}
  {D93}},\ \bibinfo {pages} {043519} (\bibinfo {year} {2016})},\ \Eprint
  {http://arxiv.org/abs/1512.01380} {arXiv:1512.01380 [hep-th]} \BibitemShut
  {NoStop}%
\bibitem [{\citenamefont {Domènech}\ \emph {et~al.}(2017)\citenamefont
  {Domènech}, \citenamefont {Hiramatsu}, \citenamefont {Lin}, \citenamefont
  {Sasaki}, \citenamefont {Shiraishi},\ and\ \citenamefont
  {Wang}}]{Domenech:2017kno}%
  \BibitemOpen
  \bibfield  {author} {\bibinfo {author} {\bibfnamefont {G.}~\bibnamefont
  {Domènech}}, \bibinfo {author} {\bibfnamefont {T.}~\bibnamefont
  {Hiramatsu}}, \bibinfo {author} {\bibfnamefont {C.}~\bibnamefont {Lin}},
  \bibinfo {author} {\bibfnamefont {M.}~\bibnamefont {Sasaki}}, \bibinfo
  {author} {\bibfnamefont {M.}~\bibnamefont {Shiraishi}}, \ and\ \bibinfo
  {author} {\bibfnamefont {Y.}~\bibnamefont {Wang}},\ }\href {\doibase
  10.1088/1475-7516/2017/05/034} {\bibfield  {journal} {\bibinfo  {journal}
  {JCAP}\ }\textbf {\bibinfo {volume} {1705}},\ \bibinfo {pages} {034}
  (\bibinfo {year} {2017})},\ \Eprint {http://arxiv.org/abs/1701.05554}
  {arXiv:1701.05554 [astro-ph.CO]} \BibitemShut {NoStop}%
\bibitem [{\citenamefont {Bartolo}\ and\ \citenamefont
  {Orlando}(2017)}]{Bartolo:2017szm}%
  \BibitemOpen
  \bibfield  {author} {\bibinfo {author} {\bibfnamefont {N.}~\bibnamefont
  {Bartolo}}\ and\ \bibinfo {author} {\bibfnamefont {G.}~\bibnamefont
  {Orlando}},\ }\href {\doibase 10.1088/1475-7516/2017/07/034} {\bibfield
  {journal} {\bibinfo  {journal} {JCAP}\ }\textbf {\bibinfo {volume} {1707}},\
  \bibinfo {pages} {034} (\bibinfo {year} {2017})},\ \Eprint
  {http://arxiv.org/abs/1706.04627} {arXiv:1706.04627 [astro-ph.CO]}
  \BibitemShut {NoStop}%
\bibitem [{\citenamefont {Naskar}\ and\ \citenamefont
  {Pal}(2018)}]{Naskar:2018rmu}%
  \BibitemOpen
  \bibfield  {author} {\bibinfo {author} {\bibfnamefont {A.}~\bibnamefont
  {Naskar}}\ and\ \bibinfo {author} {\bibfnamefont {S.}~\bibnamefont {Pal}},\
  }\href {\doibase 10.1103/PhysRevD.98.083520} {\bibfield  {journal} {\bibinfo
  {journal} {Phys. Rev.}\ }\textbf {\bibinfo {volume} {D98}},\ \bibinfo {pages}
  {083520} (\bibinfo {year} {2018})},\ \Eprint
  {http://arxiv.org/abs/1806.08178} {arXiv:1806.08178 [astro-ph.CO]}
  \BibitemShut {NoStop}%
\bibitem [{\citenamefont {Anninos}\ \emph {et~al.}(2019)\citenamefont
  {Anninos}, \citenamefont {De~Luca}, \citenamefont {Franciolini},
  \citenamefont {Kehagias},\ and\ \citenamefont {Riotto}}]{Anninos:2019nib}%
  \BibitemOpen
  \bibfield  {author} {\bibinfo {author} {\bibfnamefont {D.}~\bibnamefont
  {Anninos}}, \bibinfo {author} {\bibfnamefont {V.}~\bibnamefont {De~Luca}},
  \bibinfo {author} {\bibfnamefont {G.}~\bibnamefont {Franciolini}}, \bibinfo
  {author} {\bibfnamefont {A.}~\bibnamefont {Kehagias}}, \ and\ \bibinfo
  {author} {\bibfnamefont {A.}~\bibnamefont {Riotto}},\ }\href {\doibase
  10.1088/1475-7516/2019/04/045} {\bibfield  {journal} {\bibinfo  {journal}
  {JCAP}\ }\textbf {\bibinfo {volume} {1904}},\ \bibinfo {pages} {045}
  (\bibinfo {year} {2019})},\ \Eprint {http://arxiv.org/abs/1902.01251}
  {arXiv:1902.01251 [hep-th]} \BibitemShut {NoStop}%
\bibitem [{\citenamefont {Ozsoy}\ \emph {et~al.}(2019)\citenamefont {Ozsoy},
  \citenamefont {Mylova}, \citenamefont {Parameswaran}, \citenamefont {Powell},
  \citenamefont {Tasinato},\ and\ \citenamefont {Zavala}}]{Ozsoy:2019slf}%
  \BibitemOpen
  \bibfield  {author} {\bibinfo {author} {\bibfnamefont {O.}~\bibnamefont
  {Ozsoy}}, \bibinfo {author} {\bibfnamefont {M.}~\bibnamefont {Mylova}},
  \bibinfo {author} {\bibfnamefont {S.}~\bibnamefont {Parameswaran}}, \bibinfo
  {author} {\bibfnamefont {C.}~\bibnamefont {Powell}}, \bibinfo {author}
  {\bibfnamefont {G.}~\bibnamefont {Tasinato}}, \ and\ \bibinfo {author}
  {\bibfnamefont {I.}~\bibnamefont {Zavala}},\ }\href@noop {} {\  (\bibinfo
  {year} {2019})},\ \Eprint {http://arxiv.org/abs/1902.04976} {arXiv:1902.04976
  [hep-th]} \BibitemShut {NoStop}%
\bibitem [{\citenamefont {Shiraishi}\ \emph {et~al.}(2010)\citenamefont
  {Shiraishi}, \citenamefont {Yokoyama}, \citenamefont {Nitta}, \citenamefont
  {Ichiki},\ and\ \citenamefont {Takahashi}}]{Shiraishi:2010sm}%
  \BibitemOpen
  \bibfield  {author} {\bibinfo {author} {\bibfnamefont {M.}~\bibnamefont
  {Shiraishi}}, \bibinfo {author} {\bibfnamefont {S.}~\bibnamefont {Yokoyama}},
  \bibinfo {author} {\bibfnamefont {D.}~\bibnamefont {Nitta}}, \bibinfo
  {author} {\bibfnamefont {K.}~\bibnamefont {Ichiki}}, \ and\ \bibinfo {author}
  {\bibfnamefont {K.}~\bibnamefont {Takahashi}},\ }\href {\doibase
  10.1103/PhysRevD.82.103505} {\bibfield  {journal} {\bibinfo  {journal} {Phys.
  Rev.}\ }\textbf {\bibinfo {volume} {D82}},\ \bibinfo {pages} {103505}
  (\bibinfo {year} {2010})},\ \Eprint {http://arxiv.org/abs/1003.2096}
  {arXiv:1003.2096 [astro-ph.CO]} \BibitemShut {NoStop}%
\bibitem [{\citenamefont {Shiraishi}\ \emph
  {et~al.}(2011{\natexlab{b}})\citenamefont {Shiraishi}, \citenamefont {Nitta},
  \citenamefont {Yokoyama}, \citenamefont {Ichiki},\ and\ \citenamefont
  {Takahashi}}]{Shiraishi:2010kd}%
  \BibitemOpen
  \bibfield  {author} {\bibinfo {author} {\bibfnamefont {M.}~\bibnamefont
  {Shiraishi}}, \bibinfo {author} {\bibfnamefont {D.}~\bibnamefont {Nitta}},
  \bibinfo {author} {\bibfnamefont {S.}~\bibnamefont {Yokoyama}}, \bibinfo
  {author} {\bibfnamefont {K.}~\bibnamefont {Ichiki}}, \ and\ \bibinfo {author}
  {\bibfnamefont {K.}~\bibnamefont {Takahashi}},\ }\href {\doibase
  10.1143/PTP.125.795} {\bibfield  {journal} {\bibinfo  {journal} {Prog. Theor.
  Phys.}\ }\textbf {\bibinfo {volume} {125}},\ \bibinfo {pages} {795} (\bibinfo
  {year} {2011}{\natexlab{b}})},\ \Eprint {http://arxiv.org/abs/1012.1079}
  {arXiv:1012.1079 [astro-ph.CO]} \BibitemShut {NoStop}%
\bibitem [{\citenamefont {Shiraishi}\ \emph {et~al.}(2012)\citenamefont
  {Shiraishi}, \citenamefont {Nitta}, \citenamefont {Yokoyama},\ and\
  \citenamefont {Ichiki}}]{Shiraishi:2012rm}%
  \BibitemOpen
  \bibfield  {author} {\bibinfo {author} {\bibfnamefont {M.}~\bibnamefont
  {Shiraishi}}, \bibinfo {author} {\bibfnamefont {D.}~\bibnamefont {Nitta}},
  \bibinfo {author} {\bibfnamefont {S.}~\bibnamefont {Yokoyama}}, \ and\
  \bibinfo {author} {\bibfnamefont {K.}~\bibnamefont {Ichiki}},\ }\href
  {\doibase 10.1088/1475-7516/2012/03/041} {\bibfield  {journal} {\bibinfo
  {journal} {JCAP}\ }\textbf {\bibinfo {volume} {1203}},\ \bibinfo {pages}
  {041} (\bibinfo {year} {2012})},\ \Eprint {http://arxiv.org/abs/1201.0376}
  {arXiv:1201.0376 [astro-ph.CO]} \BibitemShut {NoStop}%
\bibitem [{\citenamefont {Shiraishi}(2012)}]{Shiraishi:2012sn}%
  \BibitemOpen
  \bibfield  {author} {\bibinfo {author} {\bibfnamefont {M.}~\bibnamefont
  {Shiraishi}},\ }\href {\doibase 10.1088/1475-7516/2012/06/015} {\bibfield
  {journal} {\bibinfo  {journal} {JCAP}\ }\textbf {\bibinfo {volume} {1206}},\
  \bibinfo {pages} {015} (\bibinfo {year} {2012})},\ \Eprint
  {http://arxiv.org/abs/1202.2847} {arXiv:1202.2847 [astro-ph.CO]} \BibitemShut
  {NoStop}%
\bibitem [{\citenamefont {Shiraishi}(2013)}]{Shiraishi:2013vha}%
  \BibitemOpen
  \bibfield  {author} {\bibinfo {author} {\bibfnamefont {M.}~\bibnamefont
  {Shiraishi}},\ }\href {\doibase 10.1088/1475-7516/2013/11/006} {\bibfield
  {journal} {\bibinfo  {journal} {JCAP}\ }\textbf {\bibinfo {volume} {1311}},\
  \bibinfo {pages} {006} (\bibinfo {year} {2013})},\ \Eprint
  {http://arxiv.org/abs/1308.2531} {arXiv:1308.2531 [astro-ph.CO]} \BibitemShut
  {NoStop}%
\bibitem [{\citenamefont {Shiraishi}\ \emph {et~al.}(2013)\citenamefont
  {Shiraishi}, \citenamefont {Ricciardone},\ and\ \citenamefont
  {Saga}}]{Shiraishi:2013kxa}%
  \BibitemOpen
  \bibfield  {author} {\bibinfo {author} {\bibfnamefont {M.}~\bibnamefont
  {Shiraishi}}, \bibinfo {author} {\bibfnamefont {A.}~\bibnamefont
  {Ricciardone}}, \ and\ \bibinfo {author} {\bibfnamefont {S.}~\bibnamefont
  {Saga}},\ }\href {\doibase 10.1088/1475-7516/2013/11/051} {\bibfield
  {journal} {\bibinfo  {journal} {JCAP}\ }\textbf {\bibinfo {volume} {1311}},\
  \bibinfo {pages} {051} (\bibinfo {year} {2013})},\ \Eprint
  {http://arxiv.org/abs/1308.6769} {arXiv:1308.6769 [astro-ph.CO]} \BibitemShut
  {NoStop}%
\bibitem [{\citenamefont {Shiraishi}\ \emph {et~al.}(2016)\citenamefont
  {Shiraishi}, \citenamefont {Hikage}, \citenamefont {Namba}, \citenamefont
  {Namikawa},\ and\ \citenamefont {Hazumi}}]{Shiraishi:2016yun}%
  \BibitemOpen
  \bibfield  {author} {\bibinfo {author} {\bibfnamefont {M.}~\bibnamefont
  {Shiraishi}}, \bibinfo {author} {\bibfnamefont {C.}~\bibnamefont {Hikage}},
  \bibinfo {author} {\bibfnamefont {R.}~\bibnamefont {Namba}}, \bibinfo
  {author} {\bibfnamefont {T.}~\bibnamefont {Namikawa}}, \ and\ \bibinfo
  {author} {\bibfnamefont {M.}~\bibnamefont {Hazumi}},\ }\href {\doibase
  10.1103/PhysRevD.94.043506} {\bibfield  {journal} {\bibinfo  {journal} {Phys.
  Rev.}\ }\textbf {\bibinfo {volume} {D94}},\ \bibinfo {pages} {043506}
  (\bibinfo {year} {2016})},\ \Eprint {http://arxiv.org/abs/1606.06082}
  {arXiv:1606.06082 [astro-ph.CO]} \BibitemShut {NoStop}%
\bibitem [{\citenamefont {Shiraishi}\ \emph
  {et~al.}(2011{\natexlab{c}})\citenamefont {Shiraishi}, \citenamefont
  {Nitta},\ and\ \citenamefont {Yokoyama}}]{Shiraishi:2011st}%
  \BibitemOpen
  \bibfield  {author} {\bibinfo {author} {\bibfnamefont {M.}~\bibnamefont
  {Shiraishi}}, \bibinfo {author} {\bibfnamefont {D.}~\bibnamefont {Nitta}}, \
  and\ \bibinfo {author} {\bibfnamefont {S.}~\bibnamefont {Yokoyama}},\ }\href
  {\doibase 10.1143/PTP.126.937} {\bibfield  {journal} {\bibinfo  {journal}
  {Prog. Theor. Phys.}\ }\textbf {\bibinfo {volume} {126}},\ \bibinfo {pages}
  {937} (\bibinfo {year} {2011}{\natexlab{c}})},\ \Eprint
  {http://arxiv.org/abs/1108.0175} {arXiv:1108.0175 [astro-ph.CO]} \BibitemShut
  {NoStop}%
\bibitem [{\citenamefont {Tahara}\ and\ \citenamefont
  {Yokoyama}(2018)}]{Tahara:2017wud}%
  \BibitemOpen
  \bibfield  {author} {\bibinfo {author} {\bibfnamefont {H.~W.~H.}\
  \bibnamefont {Tahara}}\ and\ \bibinfo {author} {\bibfnamefont
  {J.}~\bibnamefont {Yokoyama}},\ }\href {\doibase 10.1093/ptep/ptx185}
  {\bibfield  {journal} {\bibinfo  {journal} {PTEP}\ }\textbf {\bibinfo
  {volume} {2018}},\ \bibinfo {pages} {013E03} (\bibinfo {year} {2018})},\
  \Eprint {http://arxiv.org/abs/1704.08904} {arXiv:1704.08904 [astro-ph.CO]}
  \BibitemShut {NoStop}%
\bibitem [{\citenamefont {Bartolo}\ \emph {et~al.}(2019)\citenamefont
  {Bartolo}, \citenamefont {Orlando},\ and\ \citenamefont
  {Shiraishi}}]{Bartolo:2018elp}%
  \BibitemOpen
  \bibfield  {author} {\bibinfo {author} {\bibfnamefont {N.}~\bibnamefont
  {Bartolo}}, \bibinfo {author} {\bibfnamefont {G.}~\bibnamefont {Orlando}}, \
  and\ \bibinfo {author} {\bibfnamefont {M.}~\bibnamefont {Shiraishi}},\ }\href
  {\doibase 10.1088/1475-7516/2019/01/050} {\bibfield  {journal} {\bibinfo
  {journal} {JCAP}\ }\textbf {\bibinfo {volume} {1901}},\ \bibinfo {pages}
  {050} (\bibinfo {year} {2019})},\ \Eprint {http://arxiv.org/abs/1809.11170}
  {arXiv:1809.11170 [astro-ph.CO]} \BibitemShut {NoStop}%
\bibitem [{\citenamefont {Meerburg}\ \emph {et~al.}(2016)\citenamefont
  {Meerburg}, \citenamefont {Meyers}, \citenamefont {van Engelen},\ and\
  \citenamefont {Ali-Haïmoud}}]{Meerburg:2016ecv}%
  \BibitemOpen
  \bibfield  {author} {\bibinfo {author} {\bibfnamefont {P.~D.}\ \bibnamefont
  {Meerburg}}, \bibinfo {author} {\bibfnamefont {J.}~\bibnamefont {Meyers}},
  \bibinfo {author} {\bibfnamefont {A.}~\bibnamefont {van Engelen}}, \ and\
  \bibinfo {author} {\bibfnamefont {Y.}~\bibnamefont {Ali-Haïmoud}},\ }\href
  {\doibase 10.1103/PhysRevD.93.123511} {\bibfield  {journal} {\bibinfo
  {journal} {Phys. Rev.}\ }\textbf {\bibinfo {volume} {D93}},\ \bibinfo {pages}
  {123511} (\bibinfo {year} {2016})},\ \Eprint
  {http://arxiv.org/abs/1603.02243} {arXiv:1603.02243 [astro-ph.CO]}
  \BibitemShut {NoStop}%
\bibitem [{\citenamefont {Kothari}\ and\ \citenamefont
  {Nandi}(2019)}]{Kothari:2019yyw}%
  \BibitemOpen
  \bibfield  {author} {\bibinfo {author} {\bibfnamefont {R.}~\bibnamefont
  {Kothari}}\ and\ \bibinfo {author} {\bibfnamefont {D.}~\bibnamefont
  {Nandi}},\ }\href@noop {} {\  (\bibinfo {year} {2019})},\ \Eprint
  {http://arxiv.org/abs/1901.06538} {arXiv:1901.06538 [astro-ph.CO]}
  \BibitemShut {NoStop}%
\bibitem [{\citenamefont {Shiraishi}(2016)}]{Shiraishi:2016ads}%
  \BibitemOpen
  \bibfield  {author} {\bibinfo {author} {\bibfnamefont {M.}~\bibnamefont
  {Shiraishi}},\ }\href {\doibase 10.1142/S0217732316400034} {\bibfield
  {journal} {\bibinfo  {journal} {Mod. Phys. Lett.}\ }\textbf {\bibinfo
  {volume} {A31}},\ \bibinfo {pages} {1640003} (\bibinfo {year}
  {2016})}\BibitemShut {NoStop}%
\bibitem [{\citenamefont {Fergusson}\ \emph {et~al.}(2010)\citenamefont
  {Fergusson}, \citenamefont {Liguori},\ and\ \citenamefont
  {Shellard}}]{Fergusson:2009nv}%
  \BibitemOpen
  \bibfield  {author} {\bibinfo {author} {\bibfnamefont {J.~R.}\ \bibnamefont
  {Fergusson}}, \bibinfo {author} {\bibfnamefont {M.}~\bibnamefont {Liguori}},
  \ and\ \bibinfo {author} {\bibfnamefont {E.~P.~S.}\ \bibnamefont
  {Shellard}},\ }\href {\doibase 10.1103/PhysRevD.82.023502} {\bibfield
  {journal} {\bibinfo  {journal} {Phys. Rev.}\ }\textbf {\bibinfo {volume}
  {D82}},\ \bibinfo {pages} {023502} (\bibinfo {year} {2010})},\ \Eprint
  {http://arxiv.org/abs/0912.5516} {arXiv:0912.5516 [astro-ph.CO]} \BibitemShut
  {NoStop}%
\bibitem [{\citenamefont {Fergusson}\ \emph {et~al.}(2012)\citenamefont
  {Fergusson}, \citenamefont {Liguori},\ and\ \citenamefont
  {Shellard}}]{Fergusson:2010dm}%
  \BibitemOpen
  \bibfield  {author} {\bibinfo {author} {\bibfnamefont {J.~R.}\ \bibnamefont
  {Fergusson}}, \bibinfo {author} {\bibfnamefont {M.}~\bibnamefont {Liguori}},
  \ and\ \bibinfo {author} {\bibfnamefont {E.~P.~S.}\ \bibnamefont
  {Shellard}},\ }\href {\doibase 10.1088/1475-7516/2012/12/032} {\bibfield
  {journal} {\bibinfo  {journal} {JCAP}\ }\textbf {\bibinfo {volume} {1212}},\
  \bibinfo {pages} {032} (\bibinfo {year} {2012})},\ \Eprint
  {http://arxiv.org/abs/1006.1642} {arXiv:1006.1642 [astro-ph.CO]} \BibitemShut
  {NoStop}%
\bibitem [{\citenamefont {Fergusson}(2014)}]{Fergusson:2014gea}%
  \BibitemOpen
  \bibfield  {author} {\bibinfo {author} {\bibfnamefont {J.~R.}\ \bibnamefont
  {Fergusson}},\ }\href {\doibase 10.1103/PhysRevD.90.043533} {\bibfield
  {journal} {\bibinfo  {journal} {Phys. Rev.}\ }\textbf {\bibinfo {volume}
  {D90}},\ \bibinfo {pages} {043533} (\bibinfo {year} {2014})},\ \Eprint
  {http://arxiv.org/abs/1403.7949} {arXiv:1403.7949 [astro-ph.CO]} \BibitemShut
  {NoStop}%
\bibitem [{\citenamefont {Shiraishi}\ \emph {et~al.}(2019)\citenamefont
  {Shiraishi}, \citenamefont {Liguori}, \citenamefont {Fergusson},\ and\
  \citenamefont {Shellard}}]{Shiraishi:2019exr}%
  \BibitemOpen
  \bibfield  {author} {\bibinfo {author} {\bibfnamefont {M.}~\bibnamefont
  {Shiraishi}}, \bibinfo {author} {\bibfnamefont {M.}~\bibnamefont {Liguori}},
  \bibinfo {author} {\bibfnamefont {J.~R.}\ \bibnamefont {Fergusson}}, \ and\
  \bibinfo {author} {\bibfnamefont {E.~P.~S.}\ \bibnamefont {Shellard}},\
  }\href {\doibase 10.1088/1475-7516/2019/06/046} {\bibfield  {journal}
  {\bibinfo  {journal} {JCAP}\ }\textbf {\bibinfo {volume} {1906}},\ \bibinfo
  {pages} {046} (\bibinfo {year} {2019})},\ \Eprint
  {http://arxiv.org/abs/1904.02599} {arXiv:1904.02599 [astro-ph.CO]}
  \BibitemShut {NoStop}%
\bibitem [{\citenamefont {Shiraishi}\ \emph {et~al.}(2014)\citenamefont
  {Shiraishi}, \citenamefont {Liguori},\ and\ \citenamefont
  {Fergusson}}]{Shiraishi:2014roa}%
  \BibitemOpen
  \bibfield  {author} {\bibinfo {author} {\bibfnamefont {M.}~\bibnamefont
  {Shiraishi}}, \bibinfo {author} {\bibfnamefont {M.}~\bibnamefont {Liguori}},
  \ and\ \bibinfo {author} {\bibfnamefont {J.~R.}\ \bibnamefont {Fergusson}},\
  }\href {\doibase 10.1088/1475-7516/2014/05/008} {\bibfield  {journal}
  {\bibinfo  {journal} {JCAP}\ }\textbf {\bibinfo {volume} {1405}},\ \bibinfo
  {pages} {008} (\bibinfo {year} {2014})},\ \Eprint
  {http://arxiv.org/abs/1403.4222} {arXiv:1403.4222 [astro-ph.CO]} \BibitemShut
  {NoStop}%
\bibitem [{\citenamefont {Shiraishi}\ and\ \citenamefont
  {Sekiguchi}(2014)}]{Shiraishi:2013wua}%
  \BibitemOpen
  \bibfield  {author} {\bibinfo {author} {\bibfnamefont {M.}~\bibnamefont
  {Shiraishi}}\ and\ \bibinfo {author} {\bibfnamefont {T.}~\bibnamefont
  {Sekiguchi}},\ }\href {\doibase 10.1103/PhysRevD.90.103002} {\bibfield
  {journal} {\bibinfo  {journal} {Phys. Rev.}\ }\textbf {\bibinfo {volume}
  {D90}},\ \bibinfo {pages} {103002} (\bibinfo {year} {2014})},\ \Eprint
  {http://arxiv.org/abs/1304.7277} {arXiv:1304.7277 [astro-ph.CO]} \BibitemShut
  {NoStop}%
\bibitem [{\citenamefont {Shiraishi}\ \emph {et~al.}(2015)\citenamefont
  {Shiraishi}, \citenamefont {Liguori},\ and\ \citenamefont
  {Fergusson}}]{Shiraishi:2014ila}%
  \BibitemOpen
  \bibfield  {author} {\bibinfo {author} {\bibfnamefont {M.}~\bibnamefont
  {Shiraishi}}, \bibinfo {author} {\bibfnamefont {M.}~\bibnamefont {Liguori}},
  \ and\ \bibinfo {author} {\bibfnamefont {J.~R.}\ \bibnamefont {Fergusson}},\
  }\href {\doibase 10.1088/1475-7516/2015/01/007} {\bibfield  {journal}
  {\bibinfo  {journal} {JCAP}\ }\textbf {\bibinfo {volume} {1501}},\ \bibinfo
  {pages} {007} (\bibinfo {year} {2015})},\ \Eprint
  {http://arxiv.org/abs/1409.0265} {arXiv:1409.0265 [astro-ph.CO]} \BibitemShut
  {NoStop}%
\bibitem [{\citenamefont {Shiraishi}\ \emph {et~al.}(2018)\citenamefont
  {Shiraishi}, \citenamefont {Liguori},\ and\ \citenamefont
  {Fergusson}}]{Shiraishi:2017yrq}%
  \BibitemOpen
  \bibfield  {author} {\bibinfo {author} {\bibfnamefont {M.}~\bibnamefont
  {Shiraishi}}, \bibinfo {author} {\bibfnamefont {M.}~\bibnamefont {Liguori}},
  \ and\ \bibinfo {author} {\bibfnamefont {J.~R.}\ \bibnamefont {Fergusson}},\
  }\href {\doibase 10.1088/1475-7516/2018/01/016} {\bibfield  {journal}
  {\bibinfo  {journal} {JCAP}\ }\textbf {\bibinfo {volume} {1801}},\ \bibinfo
  {pages} {016} (\bibinfo {year} {2018})},\ \Eprint
  {http://arxiv.org/abs/1710.06778} {arXiv:1710.06778 [astro-ph.CO]}
  \BibitemShut {NoStop}%
\bibitem [{\citenamefont {Ade}\ \emph {et~al.}(2016{\natexlab{a}})\citenamefont
  {Ade} \emph {et~al.}}]{Ade:2015cva}%
  \BibitemOpen
  \bibfield  {author} {\bibinfo {author} {\bibfnamefont {P.~A.~R.}\
  \bibnamefont {Ade}} \emph {et~al.} (\bibinfo {collaboration} {Planck}),\
  }\href {\doibase 10.1051/0004-6361/201525821} {\bibfield  {journal} {\bibinfo
   {journal} {Astron. Astrophys.}\ }\textbf {\bibinfo {volume} {594}},\
  \bibinfo {pages} {A19} (\bibinfo {year} {2016}{\natexlab{a}})},\ \Eprint
  {http://arxiv.org/abs/1502.01594} {arXiv:1502.01594 [astro-ph.CO]}
  \BibitemShut {NoStop}%
\bibitem [{\citenamefont {Ade}\ \emph {et~al.}(2016{\natexlab{b}})\citenamefont
  {Ade} \emph {et~al.}}]{Ade:2015ava}%
  \BibitemOpen
  \bibfield  {author} {\bibinfo {author} {\bibfnamefont {P.~A.~R.}\
  \bibnamefont {Ade}} \emph {et~al.} (\bibinfo {collaboration} {Planck}),\
  }\href {\doibase 10.1051/0004-6361/201525836} {\bibfield  {journal} {\bibinfo
   {journal} {Astron. Astrophys.}\ }\textbf {\bibinfo {volume} {594}},\
  \bibinfo {pages} {A17} (\bibinfo {year} {2016}{\natexlab{b}})},\ \Eprint
  {http://arxiv.org/abs/1502.01592} {arXiv:1502.01592 [astro-ph.CO]}
  \BibitemShut {NoStop}%
\bibitem [{\citenamefont {Akrami}\ \emph {et~al.}(2019)\citenamefont {Akrami}
  \emph {et~al.}}]{Akrami:2019izv}%
  \BibitemOpen
  \bibfield  {author} {\bibinfo {author} {\bibfnamefont {Y.}~\bibnamefont
  {Akrami}} \emph {et~al.} (\bibinfo {collaboration} {Planck}),\ }\href@noop {}
  {\  (\bibinfo {year} {2019})},\ \Eprint {http://arxiv.org/abs/1905.05697}
  {arXiv:1905.05697 [astro-ph.CO]} \BibitemShut {NoStop}%
\bibitem [{\citenamefont {Hazumi}\ \emph {et~al.}(2012)\citenamefont {Hazumi}
  \emph {et~al.}}]{Hazumi:2012aa}%
  \BibitemOpen
  \bibfield  {author} {\bibinfo {author} {\bibfnamefont {M.}~\bibnamefont
  {Hazumi}} \emph {et~al.} (\bibinfo {collaboration} {LiteBIRD}),\ }\href
  {\doibase 10.1117/12.926743} {\bibfield  {journal} {\bibinfo  {journal}
  {Proc. SPIE Int. Soc. Opt. Eng.}\ }\textbf {\bibinfo {volume} {8442}},\
  \bibinfo {pages} {844219} (\bibinfo {year} {2012})}\BibitemShut {NoStop}%
\bibitem [{\citenamefont {Matsumura}\ \emph {et~al.}(2013)\citenamefont
  {Matsumura} \emph {et~al.}}]{Matsumura:2013aja}%
  \BibitemOpen
  \bibfield  {author} {\bibinfo {author} {\bibfnamefont {T.}~\bibnamefont
  {Matsumura}} \emph {et~al.},\ }\href {\doibase 10.1007/s10909-013-0996-1} {\
  (\bibinfo {year} {2013}),\ 10.1007/s10909-013-0996-1},\ \bibinfo {note} {[J.
  Low. Temp. Phys.176,733(2014)]},\ \Eprint {http://arxiv.org/abs/1311.2847}
  {arXiv:1311.2847 [astro-ph.IM]} \BibitemShut {NoStop}%
\bibitem [{\citenamefont {Matsumura}\ \emph {et~al.}(2016)\citenamefont
  {Matsumura} \emph {et~al.}}]{2016JLTP..tmp..169M}%
  \BibitemOpen
  \bibfield  {author} {\bibinfo {author} {\bibfnamefont {T.}~\bibnamefont
  {Matsumura}} \emph {et~al.},\ }\href {\doibase 10.1007/s10909-016-1542-8}
  {\bibfield  {journal} {\bibinfo  {journal} {Journal of Low Temperature
  Physics}\ } (\bibinfo {year} {2016}),\ 10.1007/s10909-016-1542-8}\BibitemShut
  {NoStop}%
\bibitem [{\citenamefont {Cook}\ and\ \citenamefont
  {Sorbo}(2013)}]{Cook:2013xea}%
  \BibitemOpen
  \bibfield  {author} {\bibinfo {author} {\bibfnamefont {J.~L.}\ \bibnamefont
  {Cook}}\ and\ \bibinfo {author} {\bibfnamefont {L.}~\bibnamefont {Sorbo}},\
  }\href {\doibase 10.1088/1475-7516/2013/11/047} {\bibfield  {journal}
  {\bibinfo  {journal} {JCAP}\ }\textbf {\bibinfo {volume} {1311}},\ \bibinfo
  {pages} {047} (\bibinfo {year} {2013})},\ \Eprint
  {http://arxiv.org/abs/1307.7077} {arXiv:1307.7077 [astro-ph.CO]} \BibitemShut
  {NoStop}%
\bibitem [{\citenamefont {Maleknejad}(2016)}]{Maleknejad:2016qjz}%
  \BibitemOpen
  \bibfield  {author} {\bibinfo {author} {\bibfnamefont {A.}~\bibnamefont
  {Maleknejad}},\ }\href {\doibase 10.1007/JHEP07(2016)104} {\bibfield
  {journal} {\bibinfo  {journal} {JHEP}\ }\textbf {\bibinfo {volume} {07}},\
  \bibinfo {pages} {104} (\bibinfo {year} {2016})},\ \Eprint
  {http://arxiv.org/abs/1604.03327} {arXiv:1604.03327 [hep-ph]} \BibitemShut
  {NoStop}%
\bibitem [{\citenamefont {Dimastrogiovanni}\ \emph {et~al.}(2017)\citenamefont
  {Dimastrogiovanni}, \citenamefont {Fasiello},\ and\ \citenamefont
  {Fujita}}]{Dimastrogiovanni:2016fuu}%
  \BibitemOpen
  \bibfield  {author} {\bibinfo {author} {\bibfnamefont {E.}~\bibnamefont
  {Dimastrogiovanni}}, \bibinfo {author} {\bibfnamefont {M.}~\bibnamefont
  {Fasiello}}, \ and\ \bibinfo {author} {\bibfnamefont {T.}~\bibnamefont
  {Fujita}},\ }\href {\doibase 10.1088/1475-7516/2017/01/019} {\bibfield
  {journal} {\bibinfo  {journal} {JCAP}\ }\textbf {\bibinfo {volume} {1701}},\
  \bibinfo {pages} {019} (\bibinfo {year} {2017})},\ \Eprint
  {http://arxiv.org/abs/1608.04216} {arXiv:1608.04216 [astro-ph.CO]}
  \BibitemShut {NoStop}%
\bibitem [{\citenamefont {Agrawal}\ \emph
  {et~al.}(2018{\natexlab{b}})\citenamefont {Agrawal}, \citenamefont {Fujita},\
  and\ \citenamefont {Komatsu}}]{Agrawal:2018mrg}%
  \BibitemOpen
  \bibfield  {author} {\bibinfo {author} {\bibfnamefont {A.}~\bibnamefont
  {Agrawal}}, \bibinfo {author} {\bibfnamefont {T.}~\bibnamefont {Fujita}}, \
  and\ \bibinfo {author} {\bibfnamefont {E.}~\bibnamefont {Komatsu}},\ }\href
  {\doibase 10.1088/1475-7516/2018/06/027} {\bibfield  {journal} {\bibinfo
  {journal} {JCAP}\ }\textbf {\bibinfo {volume} {1806}},\ \bibinfo {pages}
  {027} (\bibinfo {year} {2018}{\natexlab{b}})},\ \Eprint
  {http://arxiv.org/abs/1802.09284} {arXiv:1802.09284 [astro-ph.CO]}
  \BibitemShut {NoStop}%
\bibitem [{\citenamefont {Barnaby}\ \emph {et~al.}(2011)\citenamefont
  {Barnaby}, \citenamefont {Namba},\ and\ \citenamefont
  {Peloso}}]{Barnaby:2011vw}%
  \BibitemOpen
  \bibfield  {author} {\bibinfo {author} {\bibfnamefont {N.}~\bibnamefont
  {Barnaby}}, \bibinfo {author} {\bibfnamefont {R.}~\bibnamefont {Namba}}, \
  and\ \bibinfo {author} {\bibfnamefont {M.}~\bibnamefont {Peloso}},\ }\href
  {\doibase 10.1088/1475-7516/2011/04/009} {\bibfield  {journal} {\bibinfo
  {journal} {JCAP}\ }\textbf {\bibinfo {volume} {1104}},\ \bibinfo {pages}
  {009} (\bibinfo {year} {2011})},\ \Eprint {http://arxiv.org/abs/1102.4333}
  {arXiv:1102.4333 [astro-ph.CO]} \BibitemShut {NoStop}%
\bibitem [{\citenamefont {Widrow}(2002)}]{Widrow:2002ud}%
  \BibitemOpen
  \bibfield  {author} {\bibinfo {author} {\bibfnamefont {L.~M.}\ \bibnamefont
  {Widrow}},\ }\href {\doibase 10.1103/RevModPhys.74.775} {\bibfield  {journal}
  {\bibinfo  {journal} {Rev. Mod. Phys.}\ }\textbf {\bibinfo {volume} {74}},\
  \bibinfo {pages} {775} (\bibinfo {year} {2002})},\ \Eprint
  {http://arxiv.org/abs/astro-ph/0207240} {arXiv:astro-ph/0207240 [astro-ph]}
  \BibitemShut {NoStop}%
\bibitem [{\citenamefont {Kulsrud}\ and\ \citenamefont
  {Zweibel}(2008)}]{Kulsrud:2007an}%
  \BibitemOpen
  \bibfield  {author} {\bibinfo {author} {\bibfnamefont {R.~M.}\ \bibnamefont
  {Kulsrud}}\ and\ \bibinfo {author} {\bibfnamefont {E.~G.}\ \bibnamefont
  {Zweibel}},\ }\href {\doibase 10.1088/0034-4885/71/4/046901} {\bibfield
  {journal} {\bibinfo  {journal} {Rept. Prog. Phys.}\ }\textbf {\bibinfo
  {volume} {71}},\ \bibinfo {pages} {0046091} (\bibinfo {year} {2008})},\
  \Eprint {http://arxiv.org/abs/0707.2783} {arXiv:0707.2783 [astro-ph]}
  \BibitemShut {NoStop}%
\bibitem [{\citenamefont {Shaw}\ and\ \citenamefont
  {Lewis}(2010)}]{Shaw:2009nf}%
  \BibitemOpen
  \bibfield  {author} {\bibinfo {author} {\bibfnamefont {J.~R.}\ \bibnamefont
  {Shaw}}\ and\ \bibinfo {author} {\bibfnamefont {A.}~\bibnamefont {Lewis}},\
  }\href {\doibase 10.1103/PhysRevD.81.043517} {\bibfield  {journal} {\bibinfo
  {journal} {Phys. Rev.}\ }\textbf {\bibinfo {volume} {D81}},\ \bibinfo {pages}
  {043517} (\bibinfo {year} {2010})},\ \Eprint {http://arxiv.org/abs/0911.2714}
  {arXiv:0911.2714 [astro-ph.CO]} \BibitemShut {NoStop}%
\bibitem [{\citenamefont {Brown}\ and\ \citenamefont
  {Crittenden}(2005)}]{Brown:2005kr}%
  \BibitemOpen
  \bibfield  {author} {\bibinfo {author} {\bibfnamefont {I.}~\bibnamefont
  {Brown}}\ and\ \bibinfo {author} {\bibfnamefont {R.}~\bibnamefont
  {Crittenden}},\ }\href {\doibase 10.1103/PhysRevD.72.063002} {\bibfield
  {journal} {\bibinfo  {journal} {Phys. Rev.}\ }\textbf {\bibinfo {volume}
  {D72}},\ \bibinfo {pages} {063002} (\bibinfo {year} {2005})},\ \Eprint
  {http://arxiv.org/abs/astro-ph/0506570} {arXiv:astro-ph/0506570 [astro-ph]}
  \BibitemShut {NoStop}%
\bibitem [{\citenamefont {Lue}\ \emph {et~al.}(1999)\citenamefont {Lue},
  \citenamefont {Wang},\ and\ \citenamefont {Kamionkowski}}]{Lue:1998mq}%
  \BibitemOpen
  \bibfield  {author} {\bibinfo {author} {\bibfnamefont {A.}~\bibnamefont
  {Lue}}, \bibinfo {author} {\bibfnamefont {L.-M.}\ \bibnamefont {Wang}}, \
  and\ \bibinfo {author} {\bibfnamefont {M.}~\bibnamefont {Kamionkowski}},\
  }\href {\doibase 10.1103/PhysRevLett.83.1506} {\bibfield  {journal} {\bibinfo
   {journal} {Phys. Rev. Lett.}\ }\textbf {\bibinfo {volume} {83}},\ \bibinfo
  {pages} {1506} (\bibinfo {year} {1999})},\ \Eprint
  {http://arxiv.org/abs/astro-ph/9812088} {arXiv:astro-ph/9812088 [astro-ph]}
  \BibitemShut {NoStop}%
\bibitem [{\citenamefont {Alexander}\ and\ \citenamefont
  {Martin}(2005)}]{Alexander:2004wk}%
  \BibitemOpen
  \bibfield  {author} {\bibinfo {author} {\bibfnamefont {S.}~\bibnamefont
  {Alexander}}\ and\ \bibinfo {author} {\bibfnamefont {J.}~\bibnamefont
  {Martin}},\ }\href {\doibase 10.1103/PhysRevD.71.063526} {\bibfield
  {journal} {\bibinfo  {journal} {Phys. Rev.}\ }\textbf {\bibinfo {volume}
  {D71}},\ \bibinfo {pages} {063526} (\bibinfo {year} {2005})},\ \Eprint
  {http://arxiv.org/abs/hep-th/0410230} {arXiv:hep-th/0410230 [hep-th]}
  \BibitemShut {NoStop}%
\bibitem [{\citenamefont {Soda}\ \emph {et~al.}(2011)\citenamefont {Soda},
  \citenamefont {Kodama},\ and\ \citenamefont {Nozawa}}]{Soda:2011am}%
  \BibitemOpen
  \bibfield  {author} {\bibinfo {author} {\bibfnamefont {J.}~\bibnamefont
  {Soda}}, \bibinfo {author} {\bibfnamefont {H.}~\bibnamefont {Kodama}}, \ and\
  \bibinfo {author} {\bibfnamefont {M.}~\bibnamefont {Nozawa}},\ }\href
  {\doibase 10.1007/JHEP08(2011)067} {\bibfield  {journal} {\bibinfo  {journal}
  {JHEP}\ }\textbf {\bibinfo {volume} {08}},\ \bibinfo {pages} {067} (\bibinfo
  {year} {2011})},\ \Eprint {http://arxiv.org/abs/1106.3228} {arXiv:1106.3228
  [hep-th]} \BibitemShut {NoStop}%
\bibitem [{\citenamefont {Komatsu}\ \emph {et~al.}(2005)\citenamefont
  {Komatsu}, \citenamefont {Spergel},\ and\ \citenamefont
  {Wandelt}}]{Komatsu:2003iq}%
  \BibitemOpen
  \bibfield  {author} {\bibinfo {author} {\bibfnamefont {E.}~\bibnamefont
  {Komatsu}}, \bibinfo {author} {\bibfnamefont {D.~N.}\ \bibnamefont
  {Spergel}}, \ and\ \bibinfo {author} {\bibfnamefont {B.~D.}\ \bibnamefont
  {Wandelt}},\ }\href {\doibase 10.1086/491724} {\bibfield  {journal} {\bibinfo
   {journal} {Astrophys.J.}\ }\textbf {\bibinfo {volume} {634}},\ \bibinfo
  {pages} {14} (\bibinfo {year} {2005})},\ \Eprint
  {http://arxiv.org/abs/astro-ph/0305189} {arXiv:astro-ph/0305189 [astro-ph]}
  \BibitemShut {NoStop}%
\bibitem [{\citenamefont {Bucher}\ \emph {et~al.}(2016)\citenamefont {Bucher},
  \citenamefont {Racine},\ and\ \citenamefont {van Tent}}]{Bucher:2015ura}%
  \BibitemOpen
  \bibfield  {author} {\bibinfo {author} {\bibfnamefont {M.}~\bibnamefont
  {Bucher}}, \bibinfo {author} {\bibfnamefont {B.}~\bibnamefont {Racine}}, \
  and\ \bibinfo {author} {\bibfnamefont {B.}~\bibnamefont {van Tent}},\ }\href
  {\doibase 10.1088/1475-7516/2016/05/055} {\bibfield  {journal} {\bibinfo
  {journal} {JCAP}\ }\textbf {\bibinfo {volume} {1605}},\ \bibinfo {pages}
  {055} (\bibinfo {year} {2016})},\ \Eprint {http://arxiv.org/abs/1509.08107}
  {arXiv:1509.08107 [astro-ph.CO]} \BibitemShut {NoStop}%
\bibitem [{\citenamefont {Abazajian}\ \emph {et~al.}(2016)\citenamefont
  {Abazajian} \emph {et~al.}}]{Abazajian:2016yjj}%
  \BibitemOpen
  \bibfield  {author} {\bibinfo {author} {\bibfnamefont {K.~N.}\ \bibnamefont
  {Abazajian}} \emph {et~al.} (\bibinfo {collaboration} {CMB-S4}),\ }\href@noop
  {} {\  (\bibinfo {year} {2016})},\ \Eprint {http://arxiv.org/abs/1610.02743}
  {arXiv:1610.02743 [astro-ph.CO]} \BibitemShut {NoStop}%
\bibitem [{\citenamefont {Delabrouille}\ \emph {et~al.}(2018)\citenamefont
  {Delabrouille} \emph {et~al.}}]{Delabrouille:2017rct}%
  \BibitemOpen
  \bibfield  {author} {\bibinfo {author} {\bibfnamefont {J.}~\bibnamefont
  {Delabrouille}} \emph {et~al.} (\bibinfo {collaboration} {CORE}),\ }\href
  {\doibase 10.1088/1475-7516/2018/04/014} {\bibfield  {journal} {\bibinfo
  {journal} {JCAP}\ }\textbf {\bibinfo {volume} {1804}},\ \bibinfo {pages}
  {014} (\bibinfo {year} {2018})},\ \Eprint {http://arxiv.org/abs/1706.04516}
  {arXiv:1706.04516 [astro-ph.IM]} \BibitemShut {NoStop}%
\bibitem [{\citenamefont {Hu}(2000)}]{Hu:2000ee}%
  \BibitemOpen
  \bibfield  {author} {\bibinfo {author} {\bibfnamefont {W.}~\bibnamefont
  {Hu}},\ }\href {\doibase 10.1103/PhysRevD.62.043007} {\bibfield  {journal}
  {\bibinfo  {journal} {Phys. Rev.}\ }\textbf {\bibinfo {volume} {D62}},\
  \bibinfo {pages} {043007} (\bibinfo {year} {2000})},\ \Eprint
  {http://arxiv.org/abs/astro-ph/0001303} {arXiv:astro-ph/0001303 [astro-ph]}
  \BibitemShut {NoStop}%
\bibitem [{\citenamefont {Lewis}\ \emph {et~al.}(2011)\citenamefont {Lewis},
  \citenamefont {Challinor},\ and\ \citenamefont {Hanson}}]{Lewis:2011fk}%
  \BibitemOpen
  \bibfield  {author} {\bibinfo {author} {\bibfnamefont {A.}~\bibnamefont
  {Lewis}}, \bibinfo {author} {\bibfnamefont {A.}~\bibnamefont {Challinor}}, \
  and\ \bibinfo {author} {\bibfnamefont {D.}~\bibnamefont {Hanson}},\ }\href
  {\doibase 10.1088/1475-7516/2011/03/018} {\bibfield  {journal} {\bibinfo
  {journal} {JCAP}\ }\textbf {\bibinfo {volume} {1103}},\ \bibinfo {pages}
  {018} (\bibinfo {year} {2011})},\ \Eprint {http://arxiv.org/abs/1101.2234}
  {arXiv:1101.2234 [astro-ph.CO]} \BibitemShut {NoStop}%
\bibitem [{\citenamefont {Pratten}\ and\ \citenamefont
  {Lewis}(2016)}]{Pratten:2016dsm}%
  \BibitemOpen
  \bibfield  {author} {\bibinfo {author} {\bibfnamefont {G.}~\bibnamefont
  {Pratten}}\ and\ \bibinfo {author} {\bibfnamefont {A.}~\bibnamefont
  {Lewis}},\ }\href {\doibase 10.1088/1475-7516/2016/08/047} {\bibfield
  {journal} {\bibinfo  {journal} {JCAP}\ }\textbf {\bibinfo {volume} {1608}},\
  \bibinfo {pages} {047} (\bibinfo {year} {2016})},\ \Eprint
  {http://arxiv.org/abs/1605.05662} {arXiv:1605.05662 [astro-ph.CO]}
  \BibitemShut {NoStop}%
\bibitem [{\citenamefont {Marozzi}\ \emph {et~al.}(2016)\citenamefont
  {Marozzi}, \citenamefont {Fanizza}, \citenamefont {Di~Dio},\ and\
  \citenamefont {Durrer}}]{Marozzi:2016uob}%
  \BibitemOpen
  \bibfield  {author} {\bibinfo {author} {\bibfnamefont {G.}~\bibnamefont
  {Marozzi}}, \bibinfo {author} {\bibfnamefont {G.}~\bibnamefont {Fanizza}},
  \bibinfo {author} {\bibfnamefont {E.}~\bibnamefont {Di~Dio}}, \ and\ \bibinfo
  {author} {\bibfnamefont {R.}~\bibnamefont {Durrer}},\ }\href {\doibase
  10.1088/1475-7516/2016/09/028} {\bibfield  {journal} {\bibinfo  {journal}
  {JCAP}\ }\textbf {\bibinfo {volume} {1609}},\ \bibinfo {pages} {028}
  (\bibinfo {year} {2016})},\ \Eprint {http://arxiv.org/abs/1605.08761}
  {arXiv:1605.08761 [astro-ph.CO]} \BibitemShut {NoStop}%
\bibitem [{\citenamefont {Sunyaev}\ and\ \citenamefont
  {Zeldovich}(1972)}]{Sunyaev:1972eq}%
  \BibitemOpen
  \bibfield  {author} {\bibinfo {author} {\bibfnamefont {R.~A.}\ \bibnamefont
  {Sunyaev}}\ and\ \bibinfo {author} {\bibfnamefont {{\relax Ya}.~B.}\
  \bibnamefont {Zeldovich}},\ }\href@noop {} {\bibfield  {journal} {\bibinfo
  {journal} {Comments Astrophys. Space Phys.}\ }\textbf {\bibinfo {volume}
  {4}},\ \bibinfo {pages} {173} (\bibinfo {year} {1972})}\BibitemShut {NoStop}%
\bibitem [{\citenamefont {Jung}\ \emph {et~al.}(2018)\citenamefont {Jung},
  \citenamefont {Racine},\ and\ \citenamefont {van Tent}}]{Jung:2018rgf}%
  \BibitemOpen
  \bibfield  {author} {\bibinfo {author} {\bibfnamefont {G.}~\bibnamefont
  {Jung}}, \bibinfo {author} {\bibfnamefont {B.}~\bibnamefont {Racine}}, \ and\
  \bibinfo {author} {\bibfnamefont {B.}~\bibnamefont {van Tent}},\ }\href
  {\doibase 10.1088/1475-7516/2018/11/047} {\bibfield  {journal} {\bibinfo
  {journal} {JCAP}\ }\textbf {\bibinfo {volume} {1811}},\ \bibinfo {pages}
  {047} (\bibinfo {year} {2018})},\ \Eprint {http://arxiv.org/abs/1810.01727}
  {arXiv:1810.01727 [astro-ph.CO]} \BibitemShut {NoStop}%
\bibitem [{\citenamefont {Coulton}\ and\ \citenamefont
  {Spergel}(2019)}]{Coulton:2019bnz}%
  \BibitemOpen
  \bibfield  {author} {\bibinfo {author} {\bibfnamefont {W.~R.}\ \bibnamefont
  {Coulton}}\ and\ \bibinfo {author} {\bibfnamefont {D.~N.}\ \bibnamefont
  {Spergel}},\ }\href@noop {} {\  (\bibinfo {year} {2019})},\ \Eprint
  {http://arxiv.org/abs/1901.04515} {arXiv:1901.04515 [astro-ph.CO]}
  \BibitemShut {NoStop}%
\bibitem [{\citenamefont {Bartolo}\ \emph {et~al.}(2018)\citenamefont
  {Bartolo}, \citenamefont {Domcke}, \citenamefont {Figueroa}, \citenamefont
  {García-Bellido}, \citenamefont {Peloso}, \citenamefont {Pieroni},
  \citenamefont {Ricciardone}, \citenamefont {Sakellariadou}, \citenamefont
  {Sorbo},\ and\ \citenamefont {Tasinato}}]{Bartolo:2018qqn}%
  \BibitemOpen
  \bibfield  {author} {\bibinfo {author} {\bibfnamefont {N.}~\bibnamefont
  {Bartolo}}, \bibinfo {author} {\bibfnamefont {V.}~\bibnamefont {Domcke}},
  \bibinfo {author} {\bibfnamefont {D.~G.}\ \bibnamefont {Figueroa}}, \bibinfo
  {author} {\bibfnamefont {J.}~\bibnamefont {García-Bellido}}, \bibinfo
  {author} {\bibfnamefont {M.}~\bibnamefont {Peloso}}, \bibinfo {author}
  {\bibfnamefont {M.}~\bibnamefont {Pieroni}}, \bibinfo {author} {\bibfnamefont
  {A.}~\bibnamefont {Ricciardone}}, \bibinfo {author} {\bibfnamefont
  {M.}~\bibnamefont {Sakellariadou}}, \bibinfo {author} {\bibfnamefont
  {L.}~\bibnamefont {Sorbo}}, \ and\ \bibinfo {author} {\bibfnamefont
  {G.}~\bibnamefont {Tasinato}},\ }\href {\doibase
  10.1088/1475-7516/2018/11/034} {\bibfield  {journal} {\bibinfo  {journal}
  {JCAP}\ }\textbf {\bibinfo {volume} {1811}},\ \bibinfo {pages} {034}
  (\bibinfo {year} {2018})},\ \Eprint {http://arxiv.org/abs/1806.02819}
  {arXiv:1806.02819 [astro-ph.CO]} \BibitemShut {NoStop}%
\bibitem [{\citenamefont {Tsuneto}\ \emph {et~al.}(2019)\citenamefont
  {Tsuneto}, \citenamefont {Ito}, \citenamefont {Noumi},\ and\ \citenamefont
  {Soda}}]{Tsuneto:2018tif}%
  \BibitemOpen
  \bibfield  {author} {\bibinfo {author} {\bibfnamefont {M.}~\bibnamefont
  {Tsuneto}}, \bibinfo {author} {\bibfnamefont {A.}~\bibnamefont {Ito}},
  \bibinfo {author} {\bibfnamefont {T.}~\bibnamefont {Noumi}}, \ and\ \bibinfo
  {author} {\bibfnamefont {J.}~\bibnamefont {Soda}},\ }\href {\doibase
  10.1088/1475-7516/2019/03/032} {\bibfield  {journal} {\bibinfo  {journal}
  {JCAP}\ }\textbf {\bibinfo {volume} {1903}},\ \bibinfo {pages} {032}
  (\bibinfo {year} {2019})},\ \Eprint {http://arxiv.org/abs/1812.10615}
  {arXiv:1812.10615 [gr-qc]} \BibitemShut {NoStop}%
\bibitem [{\citenamefont {Jeong}\ and\ \citenamefont
  {Kamionkowski}(2012)}]{Jeong:2012df}%
  \BibitemOpen
  \bibfield  {author} {\bibinfo {author} {\bibfnamefont {D.}~\bibnamefont
  {Jeong}}\ and\ \bibinfo {author} {\bibfnamefont {M.}~\bibnamefont
  {Kamionkowski}},\ }\href {\doibase 10.1103/PhysRevLett.108.251301} {\bibfield
   {journal} {\bibinfo  {journal} {Phys. Rev. Lett.}\ }\textbf {\bibinfo
  {volume} {108}},\ \bibinfo {pages} {251301} (\bibinfo {year} {2012})},\
  \Eprint {http://arxiv.org/abs/1203.0302} {arXiv:1203.0302 [astro-ph.CO]}
  \BibitemShut {NoStop}%
\end{thebibliography}%


\end{document}